\magnification=\magstep1


\catcode`\@=11


\message{Loading a modification of the jyTeX macros...}

\message{modifications to plain.tex,}


\def\newcount{\alloc@0\count\countdef\insc@unt}
\def\newdimen{\alloc@1\dimen\dimendef\insc@unt}
\def\newskip{\alloc@2\skip\skipdef\insc@unt}
\def\newmuskip{\alloc@3\muskip\muskipdef\@cclvi}
\def\newtoks{\alloc@5\toks\toksdef\@cclvi}
\def\newhelp#1#2{\newtoks#1\global#1\expandafter{\csname#2\endcsname}}
\def\newread{\alloc@6\read\chardef\sixt@@n}
\def\newwrite{\alloc@7\write\chardef\sixt@@n}
\def\newfam{\alloc@8\fam\chardef\sixt@@n}
\def\newinsert#1{\global\advance\insc@unt by\m@ne
     \ch@ck0\insc@unt\count
     \ch@ck1\insc@unt\dimen
     \ch@ck2\insc@unt\skip
     \ch@ck4\insc@unt\box
     \allocationnumber=\insc@unt
     \global\chardef#1=\allocationnumber
     \wlog{\string#1=\string\insert\the\allocationnumber}}
\def\newif#1{\count@\escapechar \escapechar\m@ne
     \expandafter\expandafter\expandafter
          \xdef\@if#1{true}{\let\noexpand#1=\noexpand\iftrue}%
     \expandafter\expandafter\expandafter
          \xdef\@if#1{false}{\let\noexpand#1=\noexpand\iffalse}%
     \global\@if#1{false}\escapechar=\count@}


\newlinechar=`\^^J
\overfullrule=0pt

\message{hacks,}


\toksdef\toks@i=1
\toksdef\toks@ii=2


\def\TeX{T\kern-.1667em \lower.5ex \hbox{E}\kern-.125em X\null}
\def\jyTeX{{\leavevmode
     \raise.587ex \hbox{\it\j}\kern-.1em \lower.048ex \hbox{\it y}\kern-.12em
     \TeX}}

\let\then=\iftrue
\def\ifnoarg#1\then{\def\hack@{#1}\ifx\hack@\empty}
\def\ifundefined#1\then{%
     \expandafter\ifx\csname\expandafter\blank\string#1\endcsname\relax}
\def\useif#1\then{\csname#1\endcsname}
\def\usename#1{\csname#1\endcsname}
\def\useafter#1#2{\expandafter#1\csname#2\endcsname}

\long\def\loop#1\repeat{\def\@iterate{#1\expandafter\@iterate\fi}\@iterate
     \let\@iterate=\relax}

\let\TeXend=\end
\def\begin#1{\begingroup\def\@@blockname{#1}\usename{begin#1}}
\def\End#1{\usename{end#1}\def\hack@{#1}%
     \ifx\@@blockname\hack@
          \endgroup
     \else\err@badgroup\hack@\@@blockname
     \fi}
\def\@@blockname{}

\def\defaultoption[#1]#2{%
     \def\hack@{\ifx\hack@ii[\toks@={#2}\else\toks@={#2[#1]}\fi\the\toks@}%
     \futurelet\hack@ii\hack@}

\def\markup#1{\let\@@marksf=\empty
     \ifhmode\edef\@@marksf{\spacefactor=\the\spacefactor\relax}\/\fi
     ${}^{\hbox{\subscriptfonts#1}}$\@@marksf}


\newtoks\shortyear
\newtoks\militaryhour
\newtoks\standardhour
\newtoks\minute
\newtoks\amorpm

\def\settime{\count@=\time\divide\count@ by60
     \militaryhour=\expandafter{\number\count@}%
     {\multiply\count@ by-60 \advance\count@ by\time
          \xdef\hack@{\ifnum\count@<10 0\fi\number\count@}}%
     \minute=\expandafter{\hack@}%
     \ifnum\count@<12
          \amorpm={am}
     \else\amorpm={pm}
          \ifnum\count@>12 \advance\count@ by-12 \fi
     \fi
     \standardhour=\expandafter{\number\count@}%
     \def\hack@19##1##2{\shortyear={##1##2}}%
          \expandafter\hack@\the\year}

\def\monthword#1{%
     \ifcase#1
          $\bullet$\err@badcountervalue{monthword}%
          \or January\or February\or March\or April\or May\or June%
          \or July\or August\or September\or October\or November\or December%
     \else$\bullet$\err@badcountervalue{monthword}%
     \fi}

\def\monthabbr#1{%
     \ifcase#1
          $\bullet$\err@badcountervalue{monthabbr}%
          \or Jan\or Feb\or Mar\or Apr\or May\or Jun%
          \or Jul\or Aug\or Sep\or Oct\or Nov\or Dec%
     \else$\bullet$\err@badcountervalue{monthabbr}%
     \fi}

\def\militarytime{\the\militaryhour:\the\minute}
\def\standardtime{\the\standardhour:\the\minute}


\def\@setnumstyle#1#2{\expandafter\global\expandafter\expandafter
     \expandafter\let\expandafter\expandafter
     \csname @\expandafter\blank\string#1style\endcsname
     \csname#2\endcsname}
\def\numstyle#1{\usename{@\expandafter\blank\string#1style}#1}
\def\ifblank#1\then{\useafter\ifx{@\expandafter\blank\string#1}\blank}

\def\blank#1{}

\def\Roman#1{\expandafter\uppercase\expandafter{\romannumeral#1}}
\def\alphabetic#1{%
     \ifcase#1
          $\bullet$\err@badcountervalue{alphabetic}%
          \or a\or b\or c\or d\or e\or f\or g\or h\or i\or j\or k\or l\or m%
          \or n\or o\or p\or q\or r\or s\or t\or u\or v\or w\or x\or y\or z%
     \else$\bullet$\err@badcountervalue{alphabetic}%
     \fi}
\def\Alphabetic#1{\expandafter\uppercase\expandafter{\alphabetic{#1}}}
\def\symbols#1{%
     \ifcase#1
          $\bullet$\err@badcountervalue{symbols}%
          \or*\or\dag\or\ddag\or\S\or$\|$%
          \or**\or\dag\dag\or\ddag\ddag\or\S\S\or$\|\|$%
     \else$\bullet$\err@badcountervalue{symbols}%
     \fi}


\catcode`\^^?=13 \def^^?{\relax}

\def\trimleading#1\to#2{\edef#2{#1}%
     \expandafter\@trimleading\expandafter#2#2^^?^^?}
\def\@trimleading#1#2#3^^?{\ifx#2^^?\def#1{}\else\def#1{#2#3}\fi}

\def\trimtrailing#1\to#2{\edef#2{#1}%
     \expandafter\@trimtrailing\expandafter#2#2^^? ^^?\relax}
\def\@trimtrailing#1#2 ^^?#3{\ifx#3\relax\toks@={}%
     \else\def#1{#2}\toks@={\trimtrailing#1\to#1}\fi
     \the\toks@}

\def\trim#1\to#2{\trimleading#1\to#2\trimtrailing#2\to#2}

\catcode`\^^?=15


\long\def\additemL#1\to#2{\toks@={\^^\{#1}}\toks@ii=\expandafter{#2}%
     \xdef#2{\the\toks@\the\toks@ii}}

\long\def\additemR#1\to#2{\toks@={\^^\{#1}}\toks@ii=\expandafter{#2}%
     \xdef#2{\the\toks@ii\the\toks@}}

\def\getitemL#1\to#2{\expandafter\@getitemL#1\hack@#1#2}
\def\@getitemL\^^\#1#2\hack@#3#4{\def#4{#1}\def#3{#2}}


\newskip\headskip
\newskip\footskip

\message{document layout,}

\newif\ifdraft
\def\draft{\drafttrue\leftmargin=.5in \overfullrule=5pt }


\newskip\abovechapterskip
\newskip\belowchapterskip
\newskip\abovesectionskip
\newskip\belowsectionskip
\newskip\abovesubsectionskip
\newskip\belowsubsectionskip

\def\chapterstyle#1{\global\expandafter\let\expandafter\@chapterstyle
     \csname#1text\endcsname}
\def\sectionstyle#1{\global\expandafter\let\expandafter\@sectionstyle
     \csname#1text\endcsname}
\def\subsectionstyle#1{\global\expandafter\let\expandafter\@subsectionstyle
     \csname#1text\endcsname}

\def\CHapter#1{%
     \ifdim\lastskip=17sp \else\chapterbreak\vskip\abovechapterskip\fi
     \@chapterstyle{\ifblank\chapternumstyle\then
          \else\newchapternum=\next\chapternumformat\ \fi#1}%
     \nobreak\vskip\belowchapterskip\vskip17sp }

\def\Section#1{%
     \ifdim\lastskip=17sp \else\sectionbreak\vskip\abovesectionskip\fi
     \@sectionstyle{\ifblank\sectionnumstyle\then
          \else\newsectionnum=\next\sectionnumformat\ \fi#1}%
     \nobreak\vskip\belowsectionskip\vskip17sp }

\def\Subsection#1{%
     \ifdim\lastskip=17sp \else\subsectionbreak\vskip\abovesubsectionskip\fi
     \@subsectionstyle{\ifblank\subsectionnumstyle\then
          \else\newsubsectionnum=\next\subsectionnumformat\ \fi#1}%
     \nobreak\vskip\belowsubsectionskip\vskip17sp }


\newtoks\everybye \everybye={\par\vfil}
\outer\def\bye{\the\everybye
     \footnotecheck
     \prelabelcheck
     \streamcheck
     \supereject
     \TeXend}

\message{labels,}

\let\@@labeldef=\xdef
\newif\if@labelfile
\newwrite\@labelfile
\let\@prelabellist=\empty

\def\Label#1#2{\trim#1\to\@@labarg\edef\@@labtext{#2}%
     \edef\@@labname{lab@\@@labarg}%
     \useafter\ifundefined\@@labname\then\else\@yeslab\fi
     \useafter\@@labeldef\@@labname{#2}%
     \ifstreaming
          \expandafter\toks@\expandafter\expandafter\expandafter
               {\csname\@@labname\endcsname}%
          \immediate\write\streamout{\noexpand\Label{\@@labarg}{\the\toks@}}%
     \fi}
\def\@yeslab{%
     \useafter\ifundefined{if\@@labname}\then
          \err@labelredef\@@labarg
     \else\useif{if\@@labname}\then
               \err@labelredef\@@labarg
          \else\global\usename{\@@labname true}%
               \useafter\ifundefined{pre\@@labname}\then
               \else\useafter\ifx{pre\@@labname}\@@labtext
                    \else\err@badlabelmatch\@@labarg
                    \fi
               \fi
               \if@labelfile
               \else\global\@labelfiletrue
                    \immediate\write\sixt@@n{--> Creating file \jobname.lab}%
                    \immediate\openout\@labelfile=\jobname.lab
               \fi
               \immediate\write\@labelfile
                    {\noexpand\prelabel{\@@labarg}{\@@labtext}}%
          \fi
     \fi}

\def\putlab#1{\trim#1\to\@@labarg\edef\@@labname{lab@\@@labarg}%
     \useafter\ifundefined\@@labname\then\@nolab\else\usename\@@labname\fi}
\def\@nolab{%
     \useafter\ifundefined{pre\@@labname}\then
          \undefinedlabelformat
          \err@needlabel\@@labarg
          \useafter\xdef\@@labname{\undefinedlabelformat}%
     \else\usename{pre\@@labname}%
          \useafter\xdef\@@labname{\usename{pre\@@labname}}%
     \fi
     \useafter\newif{if\@@labname}%
     \expandafter\additemR\@@labarg\to\@prelabellist}

\def\prelabel#1{\useafter\gdef{prelab@#1}}

\def\ifundefinedlabel#1\then{%
     \expandafter\ifx\csname lab@#1\endcsname\relax}
\def\useiflab#1\then{\csname iflab@#1\endcsname}

\def\prelabelcheck{{%
     \def\^^\##1{\useiflab{##1}\then\else\err@undefinedlabel{##1}\fi}%
     \@prelabellist}}

\message{equation numbering,}

\newcount\chapternum
\newcount\sectionnum
\newcount\subsectionnum
\newcount\equationnum
\newcount\subequationnum
\newcount\figurenum
\newcount\subfigurenum
\newcount\tablenum
\newcount\subtablenum
\newcount\defnum
\newcount\subdefnum
\newcount\thmnum
\newcount\subthmnum
\newcount\lemnum
\newcount\sublemnum

\newif\if@subeqncount
\newif\if@subfigcount
\newif\if@subtblcount
\newif\if@subdefcount
\newif\if@subthmcount
\newif\if@sublemcount

\def\newchapternum{\newsectionnum=\z@\@resetnum\chapternum}
\def\newsectionnum{\newsubsectionnum=\z@\@resetnum\sectionnum}
\def\newsubsectionnum{\newequationnum=\z@\newfigurenum=\z@\newtablenum=\z@
     \newdefnum=\z@\newthmnum=\z@\newlemnum=\z@
     \@resetnum\subsectionnum}
\def\newequationnum{\newsubequationnum=\z@\@resetnum\equationnum}
\def\newsubequationnum{\@resetnum\subequationnum}
\def\newfigurenum{\newsubfigurenum=\z@\@resetnum\figurenum}
\def\newsubfigurenum{\@resetnum\subfigurenum}
\def\newtablenum{\newsubtablenum=\z@\@resetnum\tablenum}
\def\newsubtablenum{\@resetnum\subtablenum}
\def\newdefnum{\newsubdefnum=\z@\@resetnum\defnum}
\def\newsubdefnum{\@resetnum\subdefnum}
\def\newthmnum{\newsubthmnum=\z@\@resetnum\thmnum}
\def\newsubthmnum{\@resetnum\subthmnum}
\def\newlemnum{\newsublemnum=\z@\@resetnum\lemnum}
\def\newsublemnum{\@resetnum\sublemnum}

\def\@resetnum#1{\global\advance#1by1 \edef\next{\the#1\relax}\global#1}

\newchapternum=0

\def\chapternumstyle#1{\@setnumstyle\chapternum{#1}}
\def\sectionnumstyle#1{\@setnumstyle\sectionnum{#1}}
\def\subsectionnumstyle#1{\@setnumstyle\subsectionnum{#1}}
\def\equationnumstyle#1{\@setnumstyle\equationnum{#1}}
\def\subequationnumstyle#1{\@setnumstyle\subequationnum{#1}%
     \ifblank\subequationnumstyle\then\global\@subeqncountfalse\fi
     \ignorespaces}
\def\figurenumstyle#1{\@setnumstyle\figurenum{#1}}
\def\subfigurenumstyle#1{\@setnumstyle\subfigurenum{#1}%
     \ifblank\subfigurenumstyle\then\global\@subfigcountfalse\fi
     \ignorespaces}
\def\tablenumstyle#1{\@setnumstyle\tablenum{#1}}
\def\subtablenumstyle#1{\@setnumstyle\subtablenum{#1}%
     \ifblank\subtablenumstyle\then\global\@subtblcountfalse\fi
     \ignorespaces}
\def\defnumstyle#1{\@setnumstyle\defnum{#1}}
\def\subdefnumstyle#1{\@setnumstyle\subdefnum{#1}%
     \ifblank\subdefnumstyle\then\global\@subdefcountfalse\fi
     \ignorespaces}
\def\thmnumstyle#1{\@setnumstyle\thmnum{#1}}
\def\subthmnumstyle#1{\@setnumstyle\subthmnum{#1}%
     \ifblank\subthmnumstyle\then\global\@subthmcountfalse\fi
     \ignorespaces}
\def\lemnumstyle#1{\@setnumstyle\lemnum{#1}}
\def\sublemnumstyle#1{\@setnumstyle\sublemnum{#1}%
     \ifblank\sublemnumstyle\then\global\@sublemcountfalse\fi
     \ignorespaces}

\def\heqnlabel{\newequationnum=\next
          \ifblank\subequationnumstyle\then
          \else\global\@subeqncounttrue
               \newsubequationnum=\@ne
          \fi}

\def\eqnlabel#1{%
     \if@subeqncount
          \newsubequationnum=\next
     \else\heqnlabel
     \fi
     \Label{#1}{\puteqnformat}(\puteqn{#1})%
     \ifdraft\rlap{\hskip.1in{\tt#1}}\fi}

\let\puteqn=\putlab

\def\putequation#1{\useafter\ifundefined{eqn@#1}\then
     \err@undefinedeqn{#1}\else\usename{eqn@#1}\fi}

\def\eqnseriesstyle#1{\gdef\@eqnseriesstyle{#1}}
\def\begineqnseries{\subequationnumstyle{\@eqnseriesstyle}%
     \defaultoption[]\@begineqnseries}
\def\@begineqnseries[#1]{\edef\@@eqnname{#1}}
\def\endeqnseries{\subequationnumstyle{blank}%
     \expandafter\ifnoarg\@@eqnname\then
     \else\Label\@@eqnname{\puteqnformat}%
     \fi
     \aftergroup\ignorespaces}

\def\figlabel#1{%
     \if@subfigcount
          \newsubfigurenum=\next
     \else\newfigurenum=\next
          \ifblank\subfigurenumstyle\then
          \else\global\@subfigcounttrue
               \newsubfigurenum=\@ne
          \fi
     \fi
     \Label{#1}{\putfigformat}\putfig{#1}%
   }

\let\putfig=\putlab

\def\figseriesstyle#1{\gdef\@figseriesstyle{#1}}
\def\beginfigseries{\subfigurenumstyle{\@figseriesstyle}%
     \defaultoption[]\@beginfigseries}
\def\@beginfigseries[#1]{\edef\@@figname{#1}}
\def\endfigseries{\subfigurenumstyle{blank}%
     \expandafter\ifnoarg\@@figname\then
     \else\Label\@@figname{\putfigformat}%
     \fi
     \aftergroup\ignorespaces}

\def\tbllabel#1{%
     \if@subtblcount
          \newsubtablenum=\next
     \else\newtablenum=\next
          \ifblank\subtablenumstyle\then
          \else\global\@subtblcounttrue
               \newsubtablenum=\@ne
          \fi
     \fi
     \Label{#1}{\puttblformat}\puttbl{#1}%
}

\let\puttbl=\putlab

\def\tblseriesstyle#1{\gdef\@tblseriesstyle{#1}}
\def\begintblseries{\subtablenumstyle{\@tblseriesstyle}%
     \defaultoption[]\@begintblseries}
\def\@begintblseries[#1]{\edef\@@tblname{#1}}
\def\endtblseries{\subtablenumstyle{blank}%
     \expandafter\ifnoarg\@@tblname\then
     \else\Label\@@tblname{\puttblformat}%
     \fi
     \aftergroup\ignorespaces}


\def\deflab#1{%
     \if@subdefcount
          \newsubdefnum=\next
     \else\newdefnum=\next
          \ifblank\subdefnumstyle\then
          \else\global\@subdefcounttrue
               \newsubdefnum=\@ne
          \fi
     \fi
     \Label{#1}{\putdefformat}\refdef{#1}%
}

\let\refdef=\putlab

\def\defseriesstyle#1{\gdef\@defseriesstyle{#1}}
\def\begindefseries{\subtablenumstyle{\@defseriesstyle}%
     \defaultoption[]\@begindefseries}
\def\@begindefseries[#1]{\edef\@@defname{#1}}
\def\enddefseries{\subdefnumstyle{blank}%
     \expandafter\ifnoarg\@@defname\then
     \else\Label\@@defname{\putdefformat}%
     \fi
     \aftergroup\ignorespaces}

\def\thmlab#1{%
     \if@subthmcount
          \newsubthmnum=\next
     \else\newthmnum=\next
          \ifblank\subthmnumstyle\then
          \else\global\@subthmcounttrue
               \newsubthmnum=\@ne
          \fi
     \fi
     \Label{#1}{\putthmformat}\refthm{#1}%
}

\let\refthm=\putlab

\def\thmseriesstyle#1{\gdef\@thmseriesstyle{#1}}
\def\beginthmseries{\subthmnumstyle{\@thmseriesstyle}%
     \defaultoption[]\@beginthmseries}
\def\@beginthmseries[#1]{\edef\@@thmname{#1}}
\def\endthmseries{\subthmstyle{blank}%
     \expandafter\ifnoarg\@@thmname\then
     \else\Label\@@thmname{\putthmformat}%
     \fi
     \aftergroup\ignorespaces}

\def\lemlab#1{%
     \if@sublemcount
          \newsublemnum=\next
     \else\newlemnum=\next
          \ifblank\sublemnumstyle\then
          \else\global\@sublemcounttrue
               \newsublemnum=\@ne
          \fi
     \fi
     \Label{#1}{\putlemformat}\reflem{#1}%
}

\let\reflem=\putlab

\def\lemseriesstyle#1{\gdef\@lemseriesstyle{#1}}
\def\beginlemseries{\sublemnumstyle{\@lemseriesstyle}%
     \defaultoption[]\@beginlemseries}
\def\@beginlemseries[#1]{\edef\@@lemname{#1}}
\def\endlemseries{\sublemnumstyle{blank}%
     \expandafter\ifnoarg\@@lemname\then
     \else\Label\@@lemname{\putlemformat}%
     \fi
     \aftergroup\ignorespaces}

\message{reference numbering,}

\newcount\referencenum \referencenum=0
\newcount\@@prerefcount \@@prerefcount=0
\newcount\@@thisref
\newcount\@@lastref
\newcount\@@loopref
\newcount\@@refseq
\newdimen\refnumindent
\let\@undefreflist=\empty

\def\referencenumstyle#1{\@setnumstyle\referencenum{#1}}

\def\referencestyle#1{\usename{@ref#1}}

\def\@refsequential{%
     \gdef\@refpredef##1{\global\advance\referencenum by\@ne
          \let\^^\=0\Label{##1}{\^^\{\the\referencenum}}%
          \useafter\gdef{ref@\the\referencenum}{{##1}{\undefinedlabelformat}}}%
     \gdef\@reference##1##2{%
          \ifundefinedlabel##1\then
          \else\def\^^\####1{\global\@@thisref=####1\relax}\putlab{##1}%
               \useafter\gdef{ref@\the\@@thisref}{{##1}{##2}}%
          \fi}%
     \gdef\endputreferences{%
          \loop\ifnum\@@loopref<\referencenum
                    \advance\@@loopref by\@ne
                    \expandafter\expandafter\expandafter\@printreference
                         \csname ref@\the\@@loopref\endcsname
          \repeat
          \par}}

\def\@refpreordered{%
     \gdef\@refpredef##1{\global\advance\referencenum by\@ne
          \additemR##1\to\@undefreflist}%
     \gdef\@reference##1##2{%
          \ifundefinedlabel##1\then
          \else\global\advance\@@loopref by\@ne
               {\let\^^\=0\Label{##1}{\^^\{\the\@@loopref}}}%
               \@printreference{##1}{##2}%
          \fi}
     \gdef\endputreferences{%
          \def\^^\####1{\useiflab{####1}\then
               \else\reference{####1}{\undefinedlabelformat}\fi}%
          \@undefreflist
          \par}}

\def\beginprereferences{\par
     \def\reference##1##2{\global\advance\referencenum by1\@ne
          \let\^^\=0\Label{##1}{\^^\{\the\referencenum}}%
          \useafter\gdef{ref@\the\referencenum}{{##1}{##2}}}}
\def\endprereferences{\global\@@prerefcount=\the\referencenum\par}

\def\beginputreferences{\par
     \refnumindent=\z@\@@loopref=\z@
     \loop\ifnum\@@loopref<\referencenum
               \advance\@@loopref by\@ne
               \setbox\z@=\hbox{\referencenum=\@@loopref
                    \referencenumformat\enskip}%
               \ifdim\wd\z@>\refnumindent\refnumindent=\wd\z@\fi
     \repeat
     \putreferenceformat
     \@@loopref=\z@
     \loop\ifnum\@@loopref<\@@prerefcount
               \advance\@@loopref by\@ne
               \expandafter\expandafter\expandafter\@printreference
                    \csname ref@\the\@@loopref\endcsname
     \repeat
     \let\reference=\@reference}

\def\@printreference#1#2{\ifx#2\undefinedlabelformat\err@undefinedref{#1}\fi
     \noindent\ifdraft\rlap{\hskip\hsize\hskip.1in \tt#1}\fi
     \llap{\referencenum=\@@loopref\referencenumformat\enskip}#2\par}

\def\reference#1#2{{\par\refnumindent=\z@\putreferenceformat\noindent#2\par}}

\def\putref#1{\trim#1\to\@@refarg
     \expandafter\ifnoarg\@@refarg\then
          \toks@={\relax}%
     \else\@@lastref=-\@m\def\@@refsep{}\def\@more{\@nextref}%
          \toks@={\@nextref#1,,}%
     \fi\the\toks@}
\def\@nextref#1,{\trim#1\to\@@refarg
     \expandafter\ifnoarg\@@refarg\then
          \let\@more=\relax
     \else\ifundefinedlabel\@@refarg\then
               \expandafter\@refpredef\expandafter{\@@refarg}%
          \fi
          \def\^^\##1{\global\@@thisref=##1\relax}%
          \global\@@thisref=\m@ne
          \setbox\z@=\hbox{\putlab\@@refarg}%
     \fi
     \advance\@@lastref by\@ne
     \ifnum\@@lastref=\@@thisref\advance\@@refseq by\@ne\else\@@refseq=\@ne\fi
     \ifnum\@@lastref<\z@
     \else\ifnum\@@refseq<\thr@@
               \@@refsep\def\@@refsep{,}%
               \ifnum\@@lastref>\z@
                    \advance\@@lastref by\m@ne
                    {\referencenum=\@@lastref\putrefformat}%
               \else\undefinedlabelformat
               \fi
          \else\def\@@refsep{--}%
          \fi
     \fi
     \@@lastref=\@@thisref
     \@more}

\message{streaming,}

\newif\ifstreaming

\def\streamto{\defaultoption[\jobname]\@streamto}
\def\@streamto[#1]{\global\streamingtrue
     \immediate\write\sixt@@n{--> Streaming to #1.str}%
     \newwrite\streamout\immediate\openout\streamout=#1.str }

\def\streamfrom{\defaultoption[\jobname]\@streamfrom}
\def\@streamfrom[#1]{\newread\streamin\openin\streamin=#1.str
     \ifeof\streamin
          \expandafter\err@nostream\expandafter{#1.str}%
     \else\immediate\write\sixt@@n{--> Streaming from #1.str}%
          \let\@@labeldef=\gdef
          \ifstreaming
               \edef\@elc{\endlinechar=\the\endlinechar}%
               \endlinechar=\m@ne
               \loop\read\streamin to\@@scratcha
                    \ifeof\streamin
                         \streamingfalse
                    \else\toks@=\expandafter{\@@scratcha}%
                         \immediate\write\streamout{\the\toks@}%
                    \fi
                    \ifstreaming
               \repeat
               \@elc
               \input #1.str
               \streamingtrue
          \else\input #1.str
          \fi
          \let\@@labeldef=\xdef
     \fi}

\def\streamcheck{\ifstreaming
     \immediate\write\streamout{\pagenum=\the\pagenum}%
     \immediate\write\streamout{\footnotenum=\the\footnotenum}%
     \immediate\write\streamout{\referencenum=\the\referencenum}%
     \immediate\write\streamout{\chapternum=\the\chapternum}%
     \immediate\write\streamout{\sectionnum=\the\sectionnum}%
     \immediate\write\streamout{\subsectionnum=\the\subsectionnum}%
     \immediate\write\streamout{\equationnum=\the\equationnum}%
     \immediate\write\streamout{\subequationnum=\the\subequationnum}%
     \immediate\write\streamout{\figurenum=\the\figurenum}%
     \immediate\write\streamout{\subfigurenum=\the\subfigurenum}%
     \immediate\write\streamout{\tablenum=\the\tablenum}%
     \immediate\write\streamout{\subtablenum=\the\subtablenum}%
     \immediate\closeout\streamout
     \fi}


\def\err@badtypesize{%
     \errhelp={The limited availability of certain fonts requires^^J%
          that the base type size be 10pt, 12pt, or 14pt.^^J}%
     \errmessage{--> Illegal base type size}}

\def\err@badsizechange{\immediate\write\sixt@@n
     {--> Size change not allowed in math mode, ignored}}

\def\err@sizetoolarge#1{\immediate\write\sixt@@n
     {--> \noexpand#1 too big, substituting HUGE}}

\def\err@sizenotavailable#1{\immediate\write\sixt@@n
     {--> Size not available, \noexpand#1 ignored}}

\def\err@fontnotavailable#1{\immediate\write\sixt@@n
     {--> Font not available, \noexpand#1 ignored}}

\def\err@sltoit{\immediate\write\sixt@@n
     {--> Style \noexpand\sl not available, substituting \noexpand\it}%
     \it}

\def\err@bfstobf{\immediate\write\sixt@@n
     {--> Style \noexpand\bfs not available, substituting \noexpand\bf}%
     \bf}

\def\err@badgroup#1#2{%
     \errhelp={The block you have just tried to close was not the one^^J%
          most recently opened.^^J}%
     \errmessage{--> \noexpand\End{#1} doesn't match \noexpand\begin{#2}}}

\def\err@badcountervalue#1{\immediate\write\sixt@@n
     {--> Counter (#1) out of bounds}}

\def\err@extrafootnotemark{\immediate\write\sixt@@n
     {--> \noexpand\footnotemark command
          has no corresponding \noexpand\footnotetext}}

\def\err@extrafootnotetext{%
     \errhelp{You have given a \noexpand\footnotetext command without first
          specifying^^Ja \noexpand\footnotemark.^^J}%
     \errmessage{--> \noexpand\footnotetext command has no corresponding
          \noexpand\footnotemark}}

\def\err@labelredef#1{\immediate\write\sixt@@n
     {--> Label "#1" redefined}}

\def\err@badlabelmatch#1{\immediate\write\sixt@@n
     {--> Definition of label "#1" doesn't match value in \jobname.lab}}

\def\err@needlabel#1{\immediate\write\sixt@@n
     {--> Label "#1" cited before its definition}}

\def\err@undefinedlabel#1{\immediate\write\sixt@@n
     {--> Label "#1" cited but never defined}}

\def\err@undefinedeqn#1{\immediate\write\sixt@@n
     {--> Equation "#1" not defined}}

\def\err@undefinedref#1{\immediate\write\sixt@@n
     {--> Reference "#1" not defined}}

\def\err@nostream#1{%
     \errhelp={You have tried to input a stream file that doesn't exist.^^J}%
     \errmessage{--> Stream file #1 not found}}

\message{jyTeX initialization}

\everyjob{\immediate\write16{--> jyTeX version \fmtversion}%
     \edef\@@jobname{\jobname}%
     \edef\jobname{\@@jobname}%
     \settime
     \openin0=\jobname.lab
     \ifeof0
     \else\closein0
          \immediate\write16{--> Getting labels from file \jobname.lab}%
          \input\jobname.lab
     \fi}


%
     \^^\{\splittopskip}%
     \^^\{\maxdepth}%
     \^^\{\skip\topins}%
     \^^\{\skip\footins}%
     \^^\{\headskip}%
     \^^\{\footskip}}

\def\scalingskipslist{%
     \^^\{\p@renwd}%
     \^^\{\delimitershortfall}%
     \^^\{\nulldelimiterspace}%
     \^^\{\scriptspace}%
     \^^\{\jot}%
     \^^\{\normalbaselineskip}%
     \^^\{\normallineskip}%
     \^^\{\normallineskiplimit}%
     \^^\{\baselineskip}%
     \^^\{\lineskip}%
     \^^\{\lineskiplimit}%
     \^^\{\bigskipamount}%
     \^^\{\medskipamount}%
     \^^\{\smallskipamount}%
     \^^\{\parskip}%
     \^^\{\parindent}%
     \^^\{\abovedisplayskip}%
     \^^\{\belowdisplayskip}%
     \^^\{\abovedisplayshortskip}%
     \^^\{\belowdisplayshortskip}%
     \^^\{\abovechapterskip}%
     \^^\{\belowchapterskip}%
     \^^\{\abovesectionskip}%
     \^^\{\belowsectionskip}%
     \^^\{\abovesubsectionskip}%
     \^^\{\belowsubsectionskip}}


\def\twoupsetup{
     \topmargin=.75in
     \leftmargin=.5in
     \vsize=6.9in
     \hsize=4.75in
     \fullhsize=10in
     \let\draft=\relax}


\chapterstyle{left}                              
\chapternumstyle{blank}                          
\def\chapterbreak{\newpage}                      
\abovechapterskip=0pt                            
\belowchapterskip=1.5\baselineskip               
     plus.38\baselineskip minus.38\baselineskip
\def\chapternumformat{\numstyle\chapternum.}     

\sectionstyle{left}                              
\sectionnumstyle{blank}                          
\def\sectionbreak{\vskip0pt plus4\baselineskip\penalty-100
     \vskip0pt plus-4\baselineskip}              
\abovesectionskip=1.5\baselineskip               
     plus.38\baselineskip minus.38\baselineskip
\belowsectionskip=\the\baselineskip              
     plus.25\baselineskip minus.25\baselineskip
\def\sectionnumformat{
     \ifblank\chapternumstyle\then\else\numstyle\chapternum.\fi
     \numstyle\sectionnum.}

\subsectionstyle{left}                           
\subsectionnumstyle{blank}                       
\def\subsectionbreak{\vskip0pt plus4\baselineskip\penalty-100
     \vskip0pt plus-4\baselineskip}              
\abovesubsectionskip=\the\baselineskip           
     plus.25\baselineskip minus.25\baselineskip
\belowsubsectionskip=.75\baselineskip            
     plus.19\baselineskip minus.19\baselineskip
\def\subsectionnumformat{
     \ifblank\chapternumstyle\then\else\numstyle\chapternum.\fi
     \ifblank\sectionnumstyle\then\else\numstyle\sectionnum.\fi
     \numstyle\subsectionnum.}


\def\undefinedlabelformat{$\bullet$}             


\equationnumstyle{arabic}                        
\subequationnumstyle{blank}                      
\figurenumstyle{arabic}                          
\subfigurenumstyle{blank}                        
\tablenumstyle{arabic}                           
\subtablenumstyle{blank}                         
\defnumstyle{arabic}                             
\subdefnumstyle{blank}                           
\thmnumstyle{arabic}                             
\subthmnumstyle{blank}                           
\lemnumstyle{arabic}                             
\sublemnumstyle{blank}                           

\eqnseriesstyle{alphabetic}                      
\figseriesstyle{alphabetic}                      
\tblseriesstyle{alphabetic}                      
\defseriesstyle{alphabetic}                      
\thmseriesstyle{alphabetic}                      
\lemseriesstyle{alphabetic}                      

\def\puteqnformat{\hbox{
     \ifblank\chapternumstyle\then\else\numstyle\chapternum.\fi
     \ifblank\sectionnumstyle\then\else\numstyle\sectionnum.\fi
     \ifblank\subsectionnumstyle\then\else\numstyle\subsectionnum.\fi
     \numstyle\equationnum
     \numstyle\subequationnum}}
\def\putfigformat{\hbox{
     \ifblank\chapternumstyle\then\else\numstyle\chapternum.\fi
     \ifblank\sectionnumstyle\then\else\numstyle\sectionnum.\fi
     \ifblank\subsectionnumstyle\then\else\numstyle\subsectionnum.\fi
     \numstyle\figurenum
     \numstyle\subfigurenum}}
\def\puttblformat{\hbox{
     \ifblank\chapternumstyle\then\else\numstyle\chapternum.\fi
     \ifblank\sectionnumstyle\then\else\numstyle\sectionnum.\fi
     \ifblank\subsectionnumstyle\then\else\numstyle\subsectionnum.\fi
     \numstyle\tablenum
     \numstyle\subtablenum}}
\def\putdefformat{\hbox{
     \ifblank\chapternumstyle\then\else\numstyle\chapternum.\fi
     \ifblank\sectionnumstyle\then\else\numstyle\sectionnum.\fi
     \ifblank\subsectionnumstyle\then\else\numstyle\subsectionnum.\fi
     \numstyle\defnum
     \numstyle\subdefnum}}
\def\putthmformat{\hbox{
     \ifblank\chapternumstyle\then\else\numstyle\chapternum.\fi
     \ifblank\sectionnumstyle\then\else\numstyle\sectionnum.\fi
     \ifblank\subsectionnumstyle\then\else\numstyle\subsectionnum.\fi
     \numstyle\thmnum
     \numstyle\subthmnum}}
\def\putlemformat{\hbox{
     \ifblank\chapternumstyle\then\else\numstyle\chapternum.\fi
     \ifblank\sectionnumstyle\then\else\numstyle\sectionnum.\fi
     \ifblank\subsectionnumstyle\then\else\numstyle\subsectionnum.\fi
     \numstyle\lemnum
     \numstyle\sublemnum}}


\referencestyle{sequential}                      
\referencenumstyle{arabic}                       
\def\putrefformat{\numstyle\referencenum}        
\def\referencenumformat{\numstyle\referencenum.} 
\def\putreferenceformat{
     \everypar={\hangindent=1em \hangafter=1 }%
     \def\\{\hfil\break\null\hskip-1em \ignorespaces}%
     \leftskip=\refnumindent\parindent=0pt \interlinepenalty=1000 }


\def\fmtversion{2.6M (June 1992)}


\def\ref#1{(\puteqn{#1})}
\def\label#1{\eqno\eqnlabel{#1}}
\font\bigboldfont=cmbx10 scaled \magstep2
\font\boldfont=cmbx10 scaled \magstep1

\def\displayhead#1{{\bigboldfont \leftline{#1}}
\vskip-10pt
\line{\hrulefill}}
\def\section#1{\ifblank\sectionnumstyle\then
          \else\newsectionnum=\next \fi
\displayhead{\ifblank\sectionnumstyle\then\else\sectionnumformat\ \fi#1}
     }
\def\subsection#1{\advance\subsectionnum by 1
  {\boldfont \leftline{\ifblank\sectionnumstyle\then\else\sectionnumformat\fi\number\subsectionnum\
        #1}} \vskip-10pt \line{\hrulefill}}
\def\appendix#1{\ifblank\sectionnumstyle\then
          \else\newsectionnum=\next \fi
\displayhead{Appendix
    \ifblank\sectionnumstyle\then\else\sectionnumformat\ \fi#1}
     }


\catcode`\@=12

\overfullrule=0pt
\advance\vsize by 2pc
\def\cite#1{$\lbrack#1\rbrack$}
\def\bibitem#1{\parindent=9mm\item{\hbox to 7 mm{\cite{#1}\hfill}}}

\def\mn{\medskip\noindent}
\def\bn{\bigskip\noindent}
\def\OPE#1#2{#1 \circ #2}
\def\EMT{T}
\def\NOP#1#2{(#1 \, #2)}
\def\acom#1#2{[#1,#2]_{+}}
\def\W{{\cal W}}
\def\bos{\varphi}
\def\bbos{\bar{\hbox{$\bos$}}}
\def\Xp{X^{+}}
\def\Xm{X^{-}}
\def\Xpm{X^{\pm}}
\def\eps{\epsilon}
\def\der{\partial}
\def\state#1{\vert #1 \rangle}
\def\vac{\state{0}}
\def\tr{{\rm tr}}
\def\floor#1{\lfloor#1\rfloor}
\def\qb{\bar{q}}
\def\kb{\bar{k}}
\font\frak=eufm10
\font\amsmath=msbm10
\font\amsmaths=msbm8
\def\Real{\hbox{\amsmath R}}
\def\Complex{\hbox{\amsmath C}}
\def\Zed{\hbox{\amsmath Z}}
\def\zed{\hbox{\amsmaths Z}}
\setbox22 = \hbox{1}
\def\id{\rlap{1}\rlap{\kern 1pt \vbox{\hrule width 4pt depth 0 pt}}
        \rlap{\kern 3.5 pt \hbox{\vrule height \ht22 depth 0 pt}}
            \hskip\wd22}
\chapternumstyle{blank}                           
\sectionnumstyle{arabic}                          
\def\FMS{1}
\def\GuLu{2}
\def\GeTo{3}
\def\FGN{4}
\def\MoRe{5}
\def\Saleur{6}
\def\Bernard{7}
\def\MaSe{8}
\def\dFMS{9}
\def\JdBLF{10}
\def\Ras{11}
\def\Kausch{12}
\def\Gin{13}
\def\Kir{14}
\def\Borcherds{15}
\def\KaRa{16}
\def\BEHHH{17}
\def\FRRTW{18}
\def\FKW{19}
\def\dBT{20}
\def\Schell{21}
\def\CaTrZe{22}
\def\FlVa{23}
\def\KacRad{24}
\def\FKRW{25}
\def\AFMO{26}
\def\KaTo{27}
\def\Kac{28}
\def\KacRaina{29}
\def\SchS{30}
\def\FeiFr{31}
\def\Auto{32}
\def\Weyl{33}
\def\BFH{34}
\def\GinProc{35}
\def\Zam{36}
\def\BCNMcC{37}
\def\multiplet{38}
\def\EHH{39}
\def\BEHHHt{40}
\def\Wang{41}
\def\Hornf{42}
\def\KaWa{43}
\def\Gurarie{44}
\def\Flohr{45}
\def\GaKa{46}
\def\Goddard{47}
%
%
\font\large=cmbx10 scaled \magstep3
\font\bigf=cmr10 scaled \magstep2
\pageno=0
\def\folio{
\ifnum\pageno<1 \footline{\hfil} \else\number\pageno \fi}
\line{August 1997 \hfill hep-th/9708160}
\rightline{DAMTP-97-85}
\rightline{SISSA 104/97/EP}
\vskip 1.7cm
\centerline{\large Ghost Systems: A Vertex Algebra Point of View}
\vskip 1.2cm
\centerline{\bigf W.\ Eholzer\raise8pt\hbox{$1,\sharp$},
                  L.\ Feh\'er\raise8pt\hbox{$2,\flat$},
                  A.\ Honecker\raise8pt\hbox{$3,\star$}}
\vskip 0.5cm
\centerline{\it
 ${}^{1}$ DAMTP, University of Cambridge, Silver Street, Cambridge CB3 9EW, U.K.}
\centerline{\it w.eholzer@damtp.cam.ac.uk}
\vskip 0.3cm
\centerline{\it
 ${}^{2}$Department of Theoretical Physics, J\'{o}zsef Attila University,}
\centerline{\it
 Tisza Lajos krt.\ 84-86, H-6720 Szeged, Hungary}
\vskip 0.3cm
\centerline{\it
 ${}^{3}$S.I.S.S.A., Via Beirut 2-4, I-34014 Trieste, Italy}
\centerline{\it honecker@sissa.it}
\vskip 2.0cm
\centerline{\bf Abstract}
\vskip 0.2truecm
\noindent
Fermionic and bosonic ghost systems are defined each
in terms of a single vertex algebra which admits
a one-parameter family of conformal structures.
The observation that these structures are related to each other
provides a simple way to obtain character formulae for a general
twisted module of a ghost system. The $U(1)$ symmetry and its
subgroups that underly the twisted modules also define an infinite
set of invariant vertex subalgebras. Their structure 
is studied in detail from a $\W$-algebra point of view
with particular emphasis on $\Zed_N$-invariant subalgebras
of the fermionic ghost system.
\vfill
\noindent
\leftline{\hbox to 5 true cm{\hrulefill}}
\noindent
${}^{\sharp}$ 
{Supported by the EPSRC and partially by 
  PPARC and  EPSRC (grant GR/J73322).}

\noindent
${}^{\flat}$
{Supported in part by the Hungarian National Science
Fund (OTKA) under grant T016251.}

\noindent
${}^{\star}$
{Work done under support of the EC TMR Programme
{\it Integrability, non--per\-turba\-tive effects and symmetry in
Quantum Field Theories}, grant FMRX-CT96-0012.}
\eject

\noindent
\section{Introduction}
\mn
Ghost systems (originally introduced in the context of string theory --
see e.g.\ \cite{\FMS}) and their interrelation have recently
attracted renewed attention \cite{\GuLu,\GeTo,\FGN} because of
their relevance in statistical physics (see e.g.\ \cite{\MoRe,
\Saleur,\Bernard,\MaSe}). Ghost systems are also important in
conformal field theory since they serve as building blocks in
free field (Wakimoto) realizations of current algebras (see e.g.\
\cite{\dFMS,\JdBLF,\Ras} and references therein). Despite their
simplicity, these systems support remarkably rich structures on
their own as well. For example, at $c=-2$ a structure has
been exhibited \cite{\Kausch} that is quite analogous to the
$c=1$ classification \cite{\Gin,\Kir}.

The aim of the first part of this paper is to clarify the relation
between the different ghost systems and their characters.
We will advocate the usefulness of the point of view to regard all
fermionic (or bosonic) ghost systems as just one linear system which
admits a one-parameter family of inequivalent conformal structures.
The main input we use for a mathematically precise formulation
is the (twisted) Borcherds identity \cite{\Borcherds,\KaRa} which
implies in particular explicit expressions
for the action of the grading operators corresponding to the
ghost number and energy in twisted modules. Identities between
characters then follow immediately.

The second part of this paper contains a discussion of
symmetry transformations of the ghost fields leaving a chosen
energy-momentum tensor invariant and vertex sub\-algebras invariant
under such transformations.
Many (quasi-)rational conformal field theories with integer
central charge $c$ can be obtained by similar projections
of simple linear systems. Such constructions are interesting
because there is evidence \cite{\BEHHH} for the conjecture that
all rational models of (bosonic) $\W$-algebras can be obtained
by quantised Drinfeld-Sokolov reduction \cite{\FRRTW,\FKW,\dBT}
with possible exceptions at integer central charge $c$.
Examples of conformal theories which are at least
difficult to obtain by quantised Drinfeld-Sokolov reduction
are given by the classification of $c=1$ conformal field theories
\cite{\Gin,\Kir} and the classification of $c=24$ conformal field
theories with a single primary field \cite{\Schell}.
Apart from being interesting in their own right,
$\W$-algebras arising as invariant subspaces of symmetries may be
useful for physical applications, for example in the description
of the fractional quantum Hall effect \cite{\CaTrZe,\FlVa}
using quasifinite representations of $\W_{1+\infty}$
\cite{\KacRad,\FKRW,\AFMO,\KaRa,\KaTo}.

In appendix A we recall the definition of twisted modules of a vertex
algebra as well as some useful formulae which will be needed in the paper.
Appendix B contains the computation of the characters and the partition 
function of $\Zed_N$-orbifolds of the complex fermion.
\bn
\section{The fermionic ghost system}
\mn
Let us first briefly review the fermionic ghost system
$b(z)$, $c(z)$ (also called $b$-$c$ system) \cite{\FMS}
in operator-product-expansion (OPE) language before we
present a more precise definition
in the framework of vertex algebras. The $b$-$c$ system is
characterised by the following non-trivial OPE
$$
\OPE{b(z)}{c(w)} = {1 \over (z-w)} + reg.
\label{bcOPE}$$
($\OPE{b(z)}{b(w)}$ and $\OPE{c(z)}{c(w)}$ have no singular part).
Among the fields that can be built out of the fields $b(z)$ and $c(z)$,
a particularly important one is the ghost-number current
$$J(z) = \NOP{c}{b}(z) \, .
\label{bcCurrent}$$
This current satisfies the OPE
$$\OPE{J(z)}{J(w)} = {1 \over (z-w)^2} + reg. \, ,
\label{JbcOPE}$$
and its OPEs with the basic ghost fields read
$$\OPE{J(z)}{b(w)} = - {b(w) \over z-w} + reg. \, , \qquad
\OPE{J(z)}{c(w)} = {c(w) \over z-w} + reg.
\label{OPEJbc}$$
The ghost number of a field $\phi(w)$
is defined as the coefficient in the first order pole of the
OPE $\OPE{J(z)}{\phi(w)}$. In particular,
$c(z)$ is assigned the ghost number $+1$
and $b(z)$ is assigned the ghost number $-1$.
The ghost number of a normal-ordered product
is simply the sum of the ghost numbers of the constituent
fields.

The $b$-$c$ system admits a one-parameter family of
energy-momentum tensors
$$\EMT^\lambda(z)
 = (\lambda-1) \NOP{b}{\der c}(z) - \lambda \NOP{c}{\der b}(z)
 = \EMT^0(z) - \lambda \der J(z) \, .
\label{bcEMT}$$
With respect to this energy-momentum tensor,
the ghost fields become primary fields of dimension
$\dim(b) = 1-\lambda$, $\dim(c) = \lambda$. The central charge of
the associated Virasoro algebra is
$c_{\lambda} = -2 ( 6 \lambda (\lambda-1) + 1)$.

The conformal dimension of the current \ref{bcCurrent}
with respect to the energy-momentum tensor \ref{bcEMT}
is one, but it is not primary for $\lambda \ne {1 \over 2}$. 

A remark may be in place for readers more familiar with a
Lagrangian point of view. Quite often, ghost systems are
introduced by specifying an action from which one then obtains
the OPE \ref{bcOPE} and a $U(1)$-invariance of the action
gives rise to \ref{bcCurrent} as a conserved current. However,
the data encoded in an action needs to be supplemented by
specifying the transformation laws of the fundamental ghost
fields which is essentially equivalent to directly prescribing
the energy-momentum tensor \ref{bcEMT}. We have adopted an
algebraic point of view, {\it i.e.}\ prescribing OPEs in place of
an action and specifying $T_{\lambda}$ instead of transformation
laws, which is a one-to-one correspondence.

The OPE \ref{bcOPE} is invariant under the transformation
$b \mapsto {\rm e}^{-2 \pi i \alpha} b$ and
$c \mapsto {\rm e}^{ 2 \pi i \alpha} c$ which enables one to impose
twisted boundary conditions
$$\lim_{\theta \to 2 \pi} b({\rm e}^{i \theta} z)
     = {\rm e}^{-2 \pi i \alpha} b(z) \, , \qquad
  \lim_{\theta \to 2 \pi} c({\rm e}^{i \theta} z)
     = {\rm e}^{ 2 \pi i \alpha} c(z) \, .
\label{bcBoundC}$$

In order to be more precise we first expand the fields $b(z)$
and $c(z)$ into modes. Since the $b$-$c$ system admits many
Virasoro elements \ref{bcEMT} we use the mathematical mode
convention which does not refer to fixed conformal dimensions of
the fields, {\it i.e.}\ a particular choice of $L_0$. A field 
$\phi(z)$ which is a composite in $b$, $c$ and satisfies the
$\beta$-twisted boundary condition 
$\phi({\rm e}^{2 \pi i}z) = {\rm e}^{-2 \pi i \beta}\phi(z)$ 
is expanded in terms of its modes $\phi^{(\alpha)}_n$ as
$$\phi(z)=\sum_{n \in \zed+\beta} \phi^{(\alpha)}_n z^{-n-1} \, .
\label{modConv}$$
Here, we have included an upper index $\alpha$ for the modes
that characterises the twist \ref{bcBoundC} of the fundamental 
ghost fields. In case of a trivial twist $\alpha=0$ we will suppress 
this upper index.

Using standard techniques one can show that the OPE \ref{bcOPE}
gives rise to the following anti-commutation relations 
\footnote{${}^{1})$}{Note that the definition of normal-ordered
products for twisted boundary conditions is more subtle than
that of commutation relations. We will define
an appropriate normal-ordering procedure later using the
twisted Borcherds identity.}
between the modes $b^{(\alpha)}_n$, $c^{(\alpha)}_k$ of the ghost
fields with the boundary conditions \ref{bcBoundC}
$$
\acom{b^{(\alpha)}_m}{c^{(\alpha)}_k} = \delta_{k+m,-1} \, , \qquad
\acom{b^{(\alpha)}_m}{b^{(\alpha)}_n} =
\acom{c^{(\alpha)}_k}{c^{(\alpha)}_l} = 0 \, ,
\label{eqbcdMod}$$
where $m, n \in \Zed+\alpha$, $k, l \in \Zed-\alpha$.

As usual one introduces a Fock space ${\cal F}^{(\alpha)}$ for
the $b$-$c$ system by introducing a cyclic vector $\vac$ with
the property
$$b^{(\alpha)}_n \vac = 0 = c^{(\alpha)}_k \vac \, , \qquad
   \forall \, n > -1+\alpha , \, k > -1 -\alpha \, .
\label{bcFockVec}$$
A basis for the Fock space ${\cal F}^{(\alpha)}$ is given by
$$b^{(\alpha)}_{n_1} \cdots b^{(\alpha)}_{n_r}
  c^{(\alpha)}_{k_1} \cdots c^{(\alpha)}_{k_s} \vac \, \qquad
{\rm with} \quad \cases{
n_1 < n_2 < \ldots < n_r \le -1+\alpha \, , & \cr
k_1 < k_2 < \ldots < k_s \le -1-\alpha \, , & \cr
}
\label{bcFockBasis}$$
and one assigns a twist $\beta = (r-s) \alpha$ to such a state.

In the untwisted case the Fock space ${\cal F}^{(0)}$ of the fermionic
ghost system carries the structure of a simple vertex algebra
(see e.g.\ section 5.1 of \cite{\Kac}). This vertex algebra
has a $\widehat{U(1)}$ vertex subalgebra generated by the state
$$\state{J} = c_{-1} b_{-1}\vac
\label{stateJbc}$$
corresponding to the current \ref{bcCurrent}, and a one-parameter
family of conformal structures given by the conformal vectors
\footnote{${}^{2})$}{
In fact, one can show that these states are the most general
possible Virasoro elements with respect to which $b$ and $c$
are primary. To see this, one makes an ansatz for $\state{\EMT^\lambda}$
as a general linear combination in terms of the basis
vectors \ref{bcFockBasis}. Then the primarity
condition $b_{1} \state{\EMT^\lambda} \sim b_{-1} \state{0}$,
$c_{1} \state{\EMT^\lambda} \sim c_{-1} \state{0}$ implies
that the conformal vector must be of the form
$\state{\EMT^\lambda} = A b_{-1} c_{-2} \state{0} +
B c_{-1} b_{-2} \state{0}$. Finally, the coefficients $A$
and $B$ are fixed by the requirement that this vector must give
rise to a Virasoro algebra. 
}
$$\state{\EMT^\lambda} = (\lambda-1) b_{-1} c_{-2} \state{0}
     - \lambda c_{-1} b_{-2} \state{0}
\label{stateEMTbc}$$
corresponding to \ref{bcEMT}.

In the case of a non-trivial
twist $\alpha \not\in \Zed$, the Fock space ${\cal F}^{(\alpha)}$
can be viewed as a twisted module of this vertex algebra.
In particular, a twisted version \cite{\KaRa} of the Borcherds
identity holds (see (M3) in appendix A). If one wants to compute
the action of the fields $\EMT^\lambda$ and $J$ in
${\cal F}^{(\alpha)}$, one can exploit a special case of this
identity which is also reported from \cite{\KaRa} in appendix A.

In order to comply with standard notations we use conventions
slightly different from \ref{modConv} for the action of
the field $\EMT^\lambda$ in ${\cal F}^{(\alpha)}$:
$$\EMT^{\lambda}(z) = \sum_{n \in \zed} L_n^{(\lambda,\alpha)} z^{-n-2}
\, .
\label{EMTmodConv}$$
Using the twisted Borcherds identity one finds that the 
modes $J_m^{\alpha}$ of the current $J$ \ref{bcCurrent} acting in 
${\cal F}^{(\alpha)}$ are explicitly given by
$$J^{(\alpha)}_m = \sum_{k < -\alpha} c^{(\alpha)}_k b^{(\alpha)}_{m-1-k}
            - \sum_{k \ge -\alpha} b^{(\alpha)}_{m-1-k} c^{(\alpha)}_k
            + \delta_{m,0} \alpha \, ,
\label{JbcNOP}$$
while for the modes of the family of energy-momentum tensors
\ref{bcEMT} one obtains
$$\eqalign{
L_n^{(\lambda,\alpha)} =
(\lambda-1) & \left( \sum_{k < \alpha} (k-n) b^{(\alpha)}_k c^{(\alpha)}_{n-1-k}
    - \sum_{k \ge \alpha} (k-n) c^{(\alpha)}_{n-1-k} b^{(\alpha)}_k \right) \cr
- \lambda & \left( \sum_{k < -\alpha} (k-n) c^{(\alpha)}_k b^{(\alpha)}_{n-1-k}
    - \sum_{k \ge -\alpha} (k-n) b^{(\alpha)}_{n-1-k} c^{(\alpha)}_k \right) \cr
+ & \left(\lambda \alpha + { \alpha (\alpha - 1) \over 2}\right)
\delta_{n,0} \cr
}\label{EMTbcNOP}$$          
(see also \cite{\Kausch} for analogous formulae in the case
$\lambda = 0$). 
These explicit formulae permit to check our results directly.
For example, one can check by elementary manipulations 
(following e.g.\ the lines of \cite{\KacRaina}) that \ref{EMTbcNOP} yields   
indeed a representation of the Virasoro algebra with the correct central     
charge. 

Now it is straightforward to check that
$$L_n^{(\lambda,\alpha)} = L_n^{(0,\alpha)} + \lambda (n+1) J^{(\alpha)}_n
\, ,
\label{EMTbcZero}$$
in agreement with \ref{bcEMT}. The specialisation of \ref{EMTbcZero}
to $n=0$ implies immediately the following identity between the
characters for the different choices of conformal elements
\ref{stateEMTbc}
$$\chi^{(\lambda,\alpha)}(q,z)
:= \tr_{{\cal F}^{(\alpha)}}
   q^{L_0^{(\lambda,\alpha)}-{c_\lambda \over 24}} z^{J^{(\alpha)}_0}
 = q^{-{{c_\lambda - c_0} \over 24}}  \chi^{(0,\alpha)}(q,q^{\lambda}z) \, .
\label{bcCharId}$$
In order to obtain an explicit expression for the r.h.s.\ one
needs the commutation relations of $b^{(\alpha)}_n$ and $c^{(\alpha)}_k$
with $J^{(\alpha)}_0$
which encode the ghost number ($-1$ for $b$ and $+1$ for $c$)
and those with $L_0^{(0,\alpha)}$ which encodes the
conformal dimensions $\dim(b) = 1$ and $\dim(c) = 0$
(for an explicit formula compare appendix A).
With this information it is easy to see from
\ref{bcFockBasis} that the character on the r.h.s.\ of \ref{bcCharId} 
equals (see also \cite{\Kausch,\GuLu})
$$\chi^{(0,\alpha)}(q,z) = q^{{1 \over 12}} q^{{\alpha (\alpha-1) \over 2}}
  z^\alpha 
  \prod_{n=1}^{\infty} (1+z q^{n+\alpha-1}) (1 + z^{-1} q^{n-\alpha})
  \, ,
\label{bcCm2char}$$
so that the most general form of the character reads
$$ \chi^{(\lambda,\alpha)}(q,z) = 
    q^{{1 \over 12}}
    q^{{1\over2} (\lambda+\alpha)(\lambda+\alpha-1)} z^{\alpha}
    \prod_{n=1}^\infty (1+z      q^{n+(\alpha+\lambda)-1})
                       (1+z^{-1} q^{n-(\alpha+\lambda) }) \, .
\label{bcChar}
$$
Note that for $\lambda={1 \over 2}$ and $\alpha=0$ we recover the character
of the complex fermion in the Neveu-Schwarz sector, and the
$\lambda = \alpha ={1 \over 2}$ case concerns the Ramond sector. 
In the latter case there is just a small asymmetry which arises from
the asymmetry of the twist \ref{bcBoundC}:
Inspection of \ref{bcFockVec} shows that precisely one of the zero modes of
the complex fermion is represented trivially for
$\lambda = \alpha ={1 \over 2}$.

The character \ref{bcChar} depends on $\lambda$ and
$\alpha$ only via the combination $\lambda+\alpha$ (apart from
the prefactor $z^{\alpha}$). This generalises a striking
similarity of the characters for the cases $\lambda=0$ and
$\lambda={1 \over 2}$ already observed in \cite{\GuLu}. The
fact that $\lambda$ and $\alpha$ appear only in this special
combination can actually be interpreted in terms of a spectral
flow (which was introduced for the $N=2$ superconformal algebra
in \cite{\SchS}). First note that the automorphism corresponding
to the boundary conditions \ref{bcBoundC} is inner and generated
by the current \ref{bcCurrent}. Furthermore, all ${\cal F}^{(\alpha)}$
viewed as $\widehat{U(1)}$-modules are isomorphic
independently of $\alpha$ as can e.g.\ be inferred by inspection
of \ref{bcFockBasis} and \ref{OPEJbc}. The spectral flow \cite{\SchS}
then relates the characters of ${\cal F}^{(\alpha)}$ to those of
${\cal F}^{(0)}$ by subtracting $\alpha \der J(z)$ from $\EMT^\lambda(z)$
which amounts to replacing $\lambda$ by $\lambda + \alpha$
in \ref{bcEMT}.

Although the computations presented above are rather elementary,
we believe that they show quite transparently how the relation
between the $c=-2$ ghost system and the $c=1$ complex fermion presented
in \cite{\GuLu} arises as a special case of an infinite set of
identities.

Instead of viewing the $b$-$c$ system as fundamental, one could
have equally well chosen the complex fermion as the starting point.
This would lead to formulae slightly different from, but equivalent
to \ref{EMTbcZero} and \ref{bcCharId}.
\bn
\section{The bosonic ghost system}
\mn
The same construction as for the fermionic ghost system can
also be applied to a bosonic ghost system, also
called $\beta$-$\gamma$ system \cite{\FMS} (see also \cite{\FeiFr}
for some remarks related to the following discussion).
Here the only non-trivial basic OPE reads
$$
\OPE{\beta(z)}{\gamma(w)} = {1 \over (z-w)} + reg.
\label{begaOPE}$$
Like for the $b$-$c$ system one can introduce
a ghost-number current
$$J(z) = \NOP{\gamma}{\beta}(z) \, ,
\label{begaCurrent}$$
which up to a sign satisfies the same OPE as in \ref{JbcOPE},
{\it i.e.}\ $\OPE{J(z)}{J(w)} = -{1 \over (z-w)^2} + reg.$.
The OPEs of the current \ref{begaCurrent} with the basic ghost
fields read
$$\OPE{J(z)}{\beta(w)} = - {\beta(w) \over z-w} + reg. \, , \qquad
\OPE{J(z)}{\gamma(w)} = {\gamma(w) \over z-w} + reg.
\label{OPEJbega}$$
As before, these OPEs can be used to define a ghost number,
which is $+1$ for the field $\gamma(z)$ and $-1$ for the field
$\beta(z)$.

Also the $\beta$-$\gamma$ systems admits a one-parameter
family of energy-momentum tensors
$$\EMT^{\lambda}(z) = (1-\lambda) \NOP{\beta}{\der \gamma}(z)
   - \lambda \NOP{\gamma}{\der \beta}(z) 
= \EMT^{0}(z) - \lambda \der J(z) \, ,
\label{begaEMT}$$
with respect to which the fields $\beta$ and $\gamma$ are primary
fields of dimension $\dim(\beta) = 1-\lambda$ and
$\dim(\gamma) = \lambda$. The Virasoro central charge 
associated to \ref{begaEMT} is $c_{\lambda} = 2 ( 6 \lambda (\lambda-1) +1)$.

Like the field \ref{bcCurrent} the current \ref{begaCurrent} has
conformal dimension one with respect to the corresponding
energy-momentum tensor \ref{begaEMT}, but is not primary for
$\lambda \ne {1 \over 2}$. 

The OPE \ref{begaOPE} has the same $U(1)$-invariance as
the OPE \ref{bcOPE} which enables one to impose again
twisted boundary conditions
$$\lim_{\theta \to 2 \pi} \beta({\rm e}^{i \theta} z)
     = {\rm e}^{-2 \pi i \alpha} \beta(z) \, , \qquad
  \lim_{\theta \to 2 \pi} \gamma({\rm e}^{i \theta} z)
     = {\rm e}^{ 2 \pi i \alpha} \gamma(z) \, .
\label{begaBoundC}$$

One obtains the following commutation relations for the
modes for $\beta^{(\alpha)}_n$ and $\gamma^{(\alpha)}_k$ which
are introduced via the expansion \ref{modConv}:
$$
[\beta^{(\alpha)}_m,\gamma^{(\alpha)}_k] = \delta_{k+m,-1} \, , \qquad
[\beta^{(\alpha)}_m,\beta^{(\alpha)}_n] =
[\gamma^{(\alpha)}_k,\gamma^{(\alpha)}_l] = 0 \, .
\label{eqbegadMod}$$
The boundary conditions \ref{begaBoundC} translate into the
choice $m, n \in \Zed+\alpha$, $k, l \in \Zed-\alpha$ in
\ref{eqbegadMod}.

We introduce a Fock space ${\cal B}^{(\alpha)}$ for the
$\beta$-$\gamma$ system with a cyclic vector $\vac$
that satisfies
$$\beta^{(\alpha)}_n \vac = 0 = \gamma^{(\alpha)}_k \vac \, , \qquad
\forall \, n > -1+\alpha , \, k > -1 -\alpha \, .
\label{begaFockVec}$$
A basis for this space is given by
$$\beta^{(\alpha)}_{n_1} \cdots \beta^{(\alpha)}_{n_r}
  \gamma^{(\alpha)}_{k_1} \cdots \gamma^{(\alpha)}_{k_s} \vac \, \qquad
{\rm with} \quad \cases{
n_1 \le n_2 \le \ldots \le n_r \le -1+\alpha \, , & \cr
k_1 \le k_2 \le \ldots \le k_s \le -1-\alpha \, . & \cr
}
\label{begaFockBasis}$$

${\cal B}^{(0)}$ carries the structure of a simple vertex algebra.
It has a $\widehat{U(1)}$ vertex subalgebra corresponding to
\ref{begaCurrent} which is generated by the state
$$\state{J} = \gamma_{-1} \beta_{-1}\vac \, . 
\label{stateJbega}$$
The conformal structures given by \ref{begaEMT} are now encoded
in the one-parameter family of conformal vectors
\footnote{${}^{3})$}{
One can show that these are the most general conformal
vectors with respect to which the fundamental ghost
fields are primary
using the same argument as for the $b$-$c$ system.
}
$$\state{\EMT^\lambda} = (1-\lambda) \beta_{-1} \gamma_{-2} \state{0}
     - \lambda \gamma_{-1} \beta_{-2} \state{0} \, .
\label{stateEMTbega}$$
For $\alpha \not\in \Zed$, ${\cal B}^{(\alpha)}$
can again be viewed as a twisted module of this vertex algebra \cite{\KaRa}.

Similar to the fermionic ghost system, one can use the 
twisted Borcherds identity (compare appendix A and (5.7) of \cite{\KaRa})
to obtain
$$J_m^{(\alpha)} =
   \sum_{k < -\alpha} \gamma^{(\alpha)}_k \beta^{(\alpha)}_{m-1-k}
      + \sum_{k \ge -\alpha} \beta^{(\alpha)}_{m-1-k} \gamma^{(\alpha)}_k
      - \delta_{m,0} \alpha \, ,
\label{JbegaNOP}$$
while for the modes of the family of energy-momentum tensors
\ref{begaEMT} one finds
$$\eqalign{
L_n^{(\lambda,\alpha)} =
(1-\lambda) & \left(
 \sum_{k < \alpha} (k-n) \beta^{(\alpha)}_k \gamma^{(\alpha)}_{n-1-k}
   + \sum_{k \ge \alpha} (k-n) \gamma^{(\alpha)}_{n-1-k} \beta^{(\alpha)}_k
 \right) \cr
- \lambda & \left(
  \sum_{k < -\alpha} (k-n) \gamma^{(\alpha)}_k \beta^{(\alpha)}_{n-1-k}
   + \sum_{k \ge -\alpha} (k-n) \beta^{(\alpha)}_{n-1-k} \gamma^{(\alpha)}_k
 \right) \cr
+ & \left(\lambda \alpha - { \alpha (\alpha - 1) \over 2}\right)
\delta_{n,0} \, .  \cr
}\label{EMTbegaNOP}$$ 
Hence the formula \ref{EMTbcZero} extends to this case which is
also in agreement with \ref{begaEMT}.
Also \ref{bcCharId} remains valid for the $\beta$-$\gamma$ system
with the obvious modification that ${\cal F}^{(\alpha)}$ should
be replaced with ${\cal B}^{(\alpha)}$ in the definition of the
character.
At $\lambda=0$ one finds ({\it c.f.}\  eq. (2.7) in \cite{\GuLu})
$$\chi^{(0,\alpha)}(q,z) = q^{-{1 \over 12}} q^{-{\alpha (\alpha-1) \over 2}}
  z^{-\alpha} 
  \prod_{n=1}^{\infty} (1 - z      q^{n+\alpha-1})^{-1} 
                       (1 - z^{-1} q^{n-\alpha  })^{-1}
  \, .
\label{begaCm2char}$$
Therefore, the most general form of the character reads
$$\chi^{(\lambda,\alpha)}(q,z) = 
    q^{-{1\over12}}
    q^{-{1\over2} (\lambda+\alpha)(\lambda+\alpha-1)} 
    z^{-\alpha}
    \prod_{n=1}^\infty (1-z      q^{n+(\alpha+\lambda)-1})^{-1}
                       (1-z^{-1} q^{n-(\alpha+\lambda)  })^{-1} \, .
\label{begaChar}
$$
Note that for $\lambda={1 \over 2}$ and $\alpha=0$ we indeed
obtain the character of the (untwisted) $c=-1$
bosonic ghost system.

Again, apart from the prefactor $z^{-\alpha}$, the character
\ref{begaChar} depends only on $\lambda+\alpha$ which generalises
the similarity between the characters for the cases $\lambda=0$ and
$\lambda={1 \over 2}$ that was observed in \cite{\GuLu}. As was
explained in detail for the $b$-$c$ system, this can be interpreted
in terms of a spectral flow.
Of course, the choice of reference point $\lambda = 0$ is
arbitrary here as well. However, at this point one can
identify the bosonic ghost system with a subspace of a
complex boson $\bos(z)$, $\bbos(z)$ via $\gamma(z) = \bos(z)$,
$\beta(z) = \der \bbos(z)$ (in \cite{\GuLu} a more
symmetric mapping was given which though involves non-local
operations such as dividing by $\sqrt{\der}$). 
\bn
\section{Symmetries and vertex subalgebras}
\mn
This section focuses on vertex subalgebras of the ghost systems
that are invariant under a certain symmetry.
We will be interested only in subalgebras that can be interpreted
within the context of conformal field theory. Therefore, we
require the chosen energy-momentum tensor to be contained within
the invariant subspace.

The ghost systems always have a $U(1)$ symmetry.
At the symmetric point $\lambda = {1 \over 2}$ ({\it i.e.}\ $c=1$
and $c=-1$ for the fermionic and
bosonic ghost system, respectively), this symmetry is enhanced
to $Sp(2,\Complex)$ rotations of the two fundamental
ghost fields. The energy-momentum tensor and the ghost-number
current are invariant under these transformations. In addition, 
if following \cite{\Kausch} one considers the ``small'' algebras
generated by $b$ and $\der c$ or $\beta$ and $\der \gamma$
({\it i.e.}\ states containing no $c_0$ or no $\gamma_0$), one
has a further extension to $Sp(2,\Complex)$ on this subspace
for $\lambda = 0$, {\it i.e.}\ at $c=-2$ or $c=2$ respectively.
The energy-momentum tensor is invariant in this case, too, but
the ghost-number current is not even contained in this small
algebra.

A twisted representation of a vertex algebra can be
decomposed into untwisted representations of the vertex subalgebra
that is invariant under the group used for the twist (see
e.g.\ \cite{\Auto} for a detailed explanation in the
case of $\Zed_2$ and section 4A of \cite{\KaTo} for a general
discussion). Because of this, the study of vertex subalgebras
that are invariant under a given symmetry can be quite useful,
apart from being interesting in its own right.
A method for studying such subalgebras using invariant theory
in the spirit of Weyl \cite{\Weyl} has been established in \cite{\BFH}.

In this section we will investigate some $\Zed_N$- and $U(1)$-invariant
vertex subalgebras of the bosonic or fermionic ghost system, respectively.
We start by presenting the necessary elements of $\Zed_N$-invariant
theory (the cases $\Zed_2$ and $U(1)$ are discussed in detail
in \cite{\BFH}; the latter case can be considered as the limit
$N \to \infty$ of $\Zed_N$). Let $\zeta$ be a generator of $\Zed_N$
that acts on two basic fields $\Xp$ and $\Xm$ via
$$\zeta \Xp = \omega \Xp \, \qquad \qquad
\zeta \Xm = \omega^{-1} \Xm \, ,
\label{ZnAct}$$
where $\omega$ is a primitive $N$th root of unity ($\omega^N = 1$). 
Then a polynomial generating set of all invariants is given by
\begineqnseries[ZnInv]
$$\eqalignno{
U^{m,n} :&= (\der^m \Xp)(\der^n \Xm) \, ,
&\eqnlabel{U1inv} \cr
V^{k_1,\ldots,k_N} :&= (\der^{k_1} \Xp) (\der^{k_2} \Xp)
                \cdots (\der^{k_N} \Xp) \, ,
&\eqnlabel{nZnInvP} \cr
W^{k_1,\ldots,k_N} :&= (\der^{k_1} \Xm) (\der^{k_2} \Xm)
                \cdots (\der^{k_N} \Xm) \, .
&\eqnlabel{nZnInvM}
}$$
\endeqnseries
On the quantum level suitable normal-ordering is understood.

For classical, (anti-)commuting fields $\Xpm$, the ring of all
relations satisfied by the
$\Zed_N$-invariants is generated by \footnote{${}^{4})$}{
(Anti-)commutativity of the $U$, $V$ and $W$ as well as identities
between fields that differ only in their order of indices are
understood to hold as well.}
\begineqnseries[ZnRel]
$$\eqalignno{
U^{k,l} U^{m,n} - \eps^3 U^{k,n} U^{m,l} &= 0 \, ,
&\eqnlabel{U1rel} \cr
U^{m,n} V^{k_1,k_2\ldots,k_N} - \eps^3 U^{k_1,n} V^{m,k_2,\ldots,k_N} &= 0 \, ,
&\eqnlabel{pZnrel} \cr
U^{m,n} W^{k_1,k_2\ldots,k_N} - \eps U^{m,k_1} W^{n,k_2,\ldots,k_N} &= 0 \, ,
&\eqnlabel{mZnrel} \cr
V^{k_1,\ldots,k_N} W^{l_1,\ldots,l_N} -
\eps^{(n-1)!} U^{k_1,l_1} \cdots U^{k_N,l_N} &=0 \, ,
&\eqnlabel{nZnrel} \cr
}$$
\endeqnseries
where $\eps = -1$ if $\Xpm$ are fermions and $\eps = 1$ if they are
bosons.

The proof that \ref{ZnInv} and \ref{ZnRel} generate the subrings of
all invariants and all (classical) relations, respectively goes as follows.
Since the $\Zed_N$-action \ref{ZnAct} is diagonal, it suffices
to consider monomials in the basic fields $\Xpm$. Denote
the exponent of the field $\Xp$ by $r$ and the exponent of $\Xm$
by $s$. Then, a monomial is invariant under \ref{ZnAct} precisely
if $r - s$ is a multiple of $N$. From this monomial we can factor
out $\floor{r/N}$ fields of type $V$ and $\floor{s/N}$ fields of
type $W$. Now we must have $r - N \floor{r/N} = s - N \floor{s/N}$,
{\it i.e.}\ the remaining fields must occur in pairs such that we
can write them in terms of $U$'s. This proves that the fields
\ref{ZnInv} generate all invariants.

On the classical level, a simple sorting argument along the
following lines shows that \ref{ZnRel} generate all relations
among the fields \ref{ZnInv}: First one defines a basis in the
space of monomials in $U$'s, $V$'s and $W$'s ({\it i.e.}\ a
standard form of monomials), e.g.\ by the requirement that at
most $N-1$ factors of type $U$ appear and ordering the fields
as well as their indices lexicographically. Then it is not
difficult to convince
oneself that an arbitrary  monomial in the $U$'s, $V$'s and $W$'s
can be brought into this standard form using the relations
\ref{ZnRel} (and (anti-)commutativity of any two fields as well
as identities between $V$'s and $W$'s differing only in their
order of indices).

It should be mentioned that so far we have regarded a field
and its derivatives as independent. Implementing the action of
$\der$ yields the following additional relations:
\begineqnseries[actDer]
$$\eqalignno{
\der U^{m,n} &= U^{m+1,n} + U^{m,n+1} \, ,
&\eqnlabel{U1actD} \cr
\der V^{k_1,\ldots,k_N} &= V^{k_1+1,\ldots,k_N} + \ldots +
                V^{k_1,\ldots,k_N + 1} \, ,
&\eqnlabel{ZnActP} \cr
\der W^{k_1,\ldots,k_N} &= W^{k_1+1,\ldots,k_N} + \ldots +
                W^{k_1,\ldots,k_N + 1} \, .
&\eqnlabel{ZnActM} \cr
}$$
\endeqnseries

For $N=2$ we clearly recover the results of \cite{\BFH} concerning
$\Zed_2$-invariants. Considering $U(1)$ as $\lim_{N \to \infty} \Zed_N$,
the invariants $W$ and $V$ become infinite order and thus drop out.
With this interpretation we also recover the result of \cite{\BFH}
that the $U(1)$-invariants are generated by \ref{U1inv},
their relations by \ref{U1rel} and that the derivative acts as
\ref{U1actD}.

The above consideration describes the structure of the classical
case completely, a prominent feature being an infinite set of
generators satisfying infinitely many relations \cite{\BFH}.
However, our main concern here are vertex algebras, {\it i.e.}\
quantum fields. Then, the relations \ref{ZnRel} acquire
normal-ordering corrections. These should be computed explicitly
to understand the structure of invariant vertex subalgebras,
since normal-ordered relations may be used to eliminate all
but finitely many of the generators \ref{ZnInv} \cite{\BFH}.
Such computations will be performed in the sequel. It is convenient
to perform these computations in the vacuum Fock space rather
than in a field formulation -- for some useful formulae we refer
to appendix A.

First, let us look at the {\it fermionic ghost system}. We set
$\Xp = b$ and $\Xm = c$ and start with the fields
$U$ that are present for all $\Zed_N$ and $U(1)$. We can choose
$U^{0,m}$ as an independent set of generators due to the
action of the derivative \ref{U1actD}. For them the classical
relations \ref{U1rel} turn into $U^{0,m} U^{0,n} = 0$, which can
easily be understood since the fermionic field $b$ satisfies a Pauli
principle. We now compute the corrections arising when normal-ordering
this relation. With some algebra (and a few formulae from
appendix A) one finds that
$$\eqalign{
\NOP{U^{0,n}}{U^{0,m}}_{-1} \vac =& U^{0,n}_{-1} U^{0,m}_{-1} \vac \cr
=& \left(
\sum_{k < 0} b_k (\der^n c)_{-2-k} - \sum_{k \ge 0} (\der^n c)_{-2-k} b_k
\right)
m! \; b_{-1} c_{-1-m} \vac \cr
=&  m! \; n! \; (-1)^n b_{-2-n} c_{-1-m} \vac
  + {(m+n+1)!  \over m+1} c_{-2-m-n} b_{-1} \vac \cr
=& - \left\{{1 \over n+1} + {1 \over m+1}\right\} (m+n+1)! \;
      b_{-1} c_{-2-m-n} \vac + L_{-1}(\ldots) \, . \cr
}\label{bcNOPsOnVac}$$
{}From this we infer that
$$\NOP{U^{0,n}}{U^{0,m}}
= {(-1)^n \over n+1} U^{n+1,m} - {1 \over m+1} U^{0,m+n+1}
= - \left\{{1 \over n+1} + {1 \over m+1}\right\}
U^{0,m+n+1}  + \der(\ldots) \, .
\label{bcFinRes}$$
With this information we have already complete control over the
$U(1)$-invariant subspace: Eq.\ \ref{bcFinRes} shows that it is
generated by $U^{0,0}$ only
and that one can express the $U^{0,n}$ with $n > 0$
as normal-ordered products or derivatives thereof. Specialization
of this result to $\lambda = {1 \over 2}$ amounts to a rederivation
of the simplest case ($L=1$) of the truncation of $\W_{1+\infty}$ to
a $\W(1,2,\ldots,L)$ at $c=L$ \cite{\FKRW}. Here we use
(the $L=1$ special case of) the fact that $\W_{1+\infty}$
at $c=L$ is isomorphic to the $U(L)$-invariants
of $L$ copies of the complex free fermion \cite{\KaRa,\FKRW,\KaTo}.
The original proof \cite{\FKRW} of these truncations involved
character considerations and quantized Drinfeld-Sokolov reduction.
Our approach using invariant theory suggests a different proof
in the case of an arbitrary integer $L$, too.

The statement that the $U(1)$-invariant vertex subalgebra is a
$\widehat{U(1)} = \W(1)$ can also be proven using a counting argument
and in this case \ref{bcFinRes} would not be really needed. 
First note that the Jacobi triple product identity (see e.g.\
(7.27) in \cite{\GinProc}) allows one to rewrite the character
\ref{bcChar} at $\lambda={1 \over 2}$ and $\alpha = 0$ as
$$\chi^{({1 \over 2},0)}(q,z) = {1 \over \eta(q)} \sum_{n \in \zed}
        q^{{n^2 \over 2}} z^n \, .
\label{bcCharSumF}$$
Using then \ref{bcCharId} as $\chi^{(\lambda,\alpha)}(q,z) =
  q^{-{{c_\lambda - 1} \over 24}}
  \chi^{({1 \over 2},\alpha)}(q,q^{\lambda-{1 \over 2}}z)$
one concludes further that
$$\chi^{(\lambda ,0)}(q,z) = {q^{-{{c_\lambda - 1} \over 24}}
\over \eta(q)} \sum_{n \in \zed}
   q^{{n^2 \over 2} + n (\lambda-{1 \over 2}) } z^n \, .
\label{bcCharSumFgen}$$
{}From this one reads off that the coefficient of each $z^n$ is
proportional to $1/\eta(q)$. In particular, the $z$-independent part
equals $q^{-{{c_\lambda - 1} \over 24}}/\eta(q)$. This is precisely
the character of an irreducible $\widehat{U(1)}$-representation.
Furthermore, $U^{0,0}$ ($=-J$ in \ref{bcCurrent}) clearly
generates a $\widehat{U(1)}$-subalgebra of the $U(1)$-invariant vertex
subalgebra. Now equality of the characters suffices to show that this vertex
subalgebra must be equal to its $\widehat{U(1)}$-subalgebra.

Another example over which we have already full control is the
$U(1)$-invariant subspace generated by $b$ and $\der c$ at
$\lambda=0$, {\it i.e.}\ $c=-2$. Since we exclude $c$ itself
here, the invariant subring is now generated by $U^{0,n}$ with
$n \ge 1$ and using \ref{bcFinRes} one can further eliminate
those with $n \ge 3$. Thus, the invariant vertex subalgebra is
now actually generated by $U^{0,1} =- \EMT^0$ and $U^{0,2}$
which has conformal dimension three, {\it i.e.}\ we recover
Zamolodchikov's seminal $\W(2,3)$ \cite{\Zam} at $c=-2$.
A realization of $\W(2,3)$ at $c=-2$ in terms of the $b$-$c$
system has already been given some time ago \cite{\BCNMcC},
but the identification with the $U(1)$-invariant subspace is
new.

Now we turn to $\Zed_N$-invariant vertex subalgebras of the
$b$-$c$ system. Then we have to include the generators
\ref{nZnInvP} and \ref{nZnInvM} and study the normal-ordered
versions of the relations \ref{pZnrel} and \ref{mZnrel}. The
first non-vanishing new generators are $V^{0,\ldots,N-1}$
and $W^{0,\ldots,N-1}$. It is straightforward to check by
elementary manipulations that
$$U^{0,0}_n V^{0,\ldots,N-1}_{-1} \vac =
U^{0,0}_n W^{0,\ldots,N-1}_{-1} \vac = 0 \qquad\qquad \forall n > 0 \, .
\label{VWhwCond}$$
This means that both $V^{0,\ldots,N-1}_{-1} \vac$ and
$W^{0,\ldots,N-1}_{-1} \vac$ are highest weight vectors for
the $\widehat{U(1)}$-algebra generated by $U^{0,0}$
(the highest weight property can also be inferred from 
\ref{bcCharSumFgen} since the character shows that there
are no states whose conformal weight and ghost number are
equal to those of the expressions in \ref{VWhwCond}).

Now we can again apply the counting argument that we already
used before: The character of an irreducible $\widehat{U(1)}$-module
equals $1/\eta(q)$ which is proportional to the coefficient
of $z^{\pm N}$ in \ref{bcCharSumFgen}. So, we can obtain all
fields with ghost number $-N$ and $N$ by iterative normal-ordered
products of $U^{0,0}$ and its derivatives
with $V^{0,\ldots,N-1}$ and $W^{0,\ldots,N-1}$,
respectively. This includes in particular all the fields
\ref{nZnInvP} and \ref{nZnInvM}. So, we have shown that the
$\Zed_N$-invariant vertex subalgebra is generated by $U^{0,0}$,
$V^{0,\ldots,N-1}$ and $W^{0,\ldots,N-1}$ which is an algebra
of type $\W(1,{N^2 \over 2} + (\lambda - {1 \over 2}) N,
{N^2 \over 2} + ({1 \over 2} - \lambda) N)$ for generic
$\lambda$. At $c=1$ we obtain a $\W(1,{N^2 \over 2},{N^2 \over 2})$
which we believe to be the same as the algebra $\hbox{\frak A}(N^2)$
constructed in Example 6.1 of \cite{\KaTo} by other methods.
Characters of this algebra are given in appendix B where we
also establish a relation to a free boson compactified at radius
$r={N \over 2}$.

For the $\Zed_N$-invariants at $c=-2$ ($\lambda = 0$) built out
of $b$ and $\der c$ we expect that the invariants are
generated by $U^{0,1}$, $U^{0,2}$, $V^{0,\ldots,N-1}$  
and $W^{1,\ldots,N}$ although we have in general no proof
of this. In other words: we conjecture the $\Zed_N$-invariant
vertex subalgebra of $b$ and $\der c$ to be of type
$\W(2,3,{N (N+1) \over 2}, {N (N+1) \over 2})$. We are now
going to prove this at least for $N=2$ where we find
a $\W(2,3^3)$ at $c=-2$ -- an algebra which was constructed
in \cite{\multiplet}. Note that a construction of $\W(2,3^3)$
in terms of the $b$-$c$ system was already given in \cite{\Kausch}
though without a proof of being identical to any invariant
vertex subalgebra. In this case we can select $V^{0,m}$ and
$W^{1,m}$ as generators due to \ref{ZnActP} and \ref{ZnActM}.
A straightforward computation shows
that
$$\eqalign{
\NOP{U^{0,1}}{V^{0,m}} &= - {m+4 \over 2 (m+2)} V^{0,m+2}
                + \der V^{0,m+1} - {1 \over 2} \der^2 V^{0,m} \, , \cr
\NOP{U^{0,1}}{W^{1,m}} &= - {m+3 \over 2 (m+1)} W^{1,m+2}
                + \der W^{1,m+1} - {1 \over 2} \der^2 W^{1,m} \, . \cr
}\label{cm2Z2rel}$$
Using these relations one can eliminate $V^{0,m}$ for $m \ge 1$
and $W^{1,m}$ for $m \ge 2$. Therefore, the $\Zed_2$-invariant
vertex subalgebra of $b$ and $\der c$ at $c=-2$ is generated by
$U^{0,1}$, $U^{0,2}$, $V^{0,1}$ and $W^{1,2}$ and thus of type
$\W(2,3^3)$.

We conclude the discussion of the $b$-$c$ system by defining
an action of $Sp(2,\Complex)$ on the subspace $\W$ generated by the
two fields $b$ and $\der c$  at $\lambda=0$, {\it i.e.}\ $c=-2$
following \cite{\Kausch}. For $G\in Sp(2,\Complex)$ this action
is simply given by $(b, \der c)^t \mapsto G(b, \der c)^t$.
This action leaves the energy-momentum tensor \ref{bcEMT}
invariant. Special subgroups of this $Sp(2,\Complex)$ are the 
groups $U(1)$ (which we have already studied in some detail)
and $\Zed_4$ acting by the following generators $X(a)$ and
${\cal J}$, respectively 
$$\eqalign{
  X(a) b &= a \;  b \, ,\quad X(a) \der c = a^{-1} \; \der c\, , \cr
  {\cal J} b &= \der c \,,\ \qquad {\cal J}\der c = - b
 \, .  \cr
}\label{GROUPSbega}$$ 
The reason for explicitly writing down the action of
the semidirect product $U(1)\times_s \Zed_4$ on $\W$
is given by the first inclusion in the sequence
$$\W(2,10) \subset \W(2,3) \subset \W(2,3^3) \subset \W \, ,
\label{W23bega}$$
which summarizes some of our results for $c=-2$. Indeed, we have
seen above that the subalgebra of $\W$ which is
invariant under $\Zed_2 =  <X(-1)> = <{\cal J}^2>$ is
the algebra of type $\W(2,3^3)$ constructed in \cite{\multiplet}
by direct methods. Primary generators for it have been constructed
in terms of the ghost system \cite{\Kausch} and are explicitly
given by eqs.\ (39)-(41) loc.cit. 
We have also seen that the $U(1)$-invariant subalgebra is
Zamolodchikov's $\W(2,3)$ \cite{\Zam} at $c=-2$. 
Its primary spin three generator is the same
linear combination of $U^{0,2}$ and $\der U^{0,1}$ as the
generator with zero ghost number among the ones of $\W(2,3^3)$
(eq.\ (40) of \cite{\Kausch}). Now, let us look at
the $U(1)\times_s \Zed_4$ invariant subalgebra. One can first
project to the $U(1)$-invariant subalgebra $\W(2,3)$ and
is then left with a $\Zed_2$ whose generator is the generator
${\cal J}$ of the original $\Zed_4$. ${\cal J}$ acts on
the primary spin three generator of this $\W(2,3)$ just by
a flip of sign. Thus the $U(1)\times_s \Zed_4$-invariant
subspace is the $\Zed_2$-orbifold of $\W(2,3)$ at $c=-2$ which was
discussed in section 2.2.2 of \cite{\BEHHH} and argued to
be the $\W$-algebra of type $\W(2,10)$ constructed
by direct methods in \cite{\EHH}. So far, it has actually
only been proven that the $U(1)\times_s \Zed_4$-invariant
algebra contains this $\W(2,10)$ as its subalgebra, neither
do we prove equality here. In any case,
the inclusions indicated by \ref{W23bega} at $c=-2$ have
been established rigorously.

We conclude this section with a look at the $U(1)$-invariant
vertex subalgebra of the {\it bosonic ghost system}.
We now set $\Xp = \beta$ and $\Xm = \gamma$ and then can choose
the $U^{0,n}$ as generators among \ref{U1inv}. Now one has to
compute the effect of normal-ordering on the relations \ref{U1rel}.
A computation along the same lines as \ref{bcNOPsOnVac} leads to
$$\eqalign{
\NOP{U^{m,n}}{U^{i,j}} &- \NOP{U^{m,j}}{U^{i,n}} =
{(-1)^{n+1} \over i+n+1} U^{i+m+n+1,j}
-{(-1)^{j+1} \over i+j+1} U^{i+m+j+1,n} \cr
&\qquad\qquad\qquad\qquad
+(-1)^m \left\{{1 \over m+j+1} - {1 \over m+n+1}\right\} U^{i,m+j+n+1} \cr
=& (-1)^{i+m} \left\{{1 \over n+i+1} + {1 \over m+j+1}
       - {1 \over i+j+1} - {1 \over m+n+1}\right\}
U^{0,m+n+i+j+1}  \cr
&+ \der(\ldots) \, . \cr
}\label{begaFinRes}$$
This is analogous to \ref{bcFinRes} for the $b$-$c$ system although
here we have a two-term relation, since there is no Pauli principle.

The coefficient of $U^{0,m+n+i+j+1}$ in \ref{begaFinRes} vanishes
if and only if $i=m$ or $n=j$. As a simple consequence we infer that
all $U^{0,l}$ with $l \ge 3$ can be eliminated using \ref{begaFinRes}
and thus the $U(1)$-invariant vertex subalgebra of the $\beta$-$\gamma$
system is an algebra of type $\W(1,2,3)$ generated by $U^{0,0}$,
$U^{0,1}$ and $U^{0,2}$. This is valid for all $\lambda$, thus
also for $c=-1$ ($\lambda = {1 \over 2}$). Using the construction of
\cite{\KaRa} we have thus rigorously proven the conjecture of
\cite{\BEHHHt} that $\W_{1 + \infty}$ truncates to a $\W(1,2,3)$ at
$c=-1$. We should mention that a proof of this fact which is also
based on the construction of \cite{\KaRa} has very
recently been worked out independently in \cite{\Wang}. Actually,
a slightly stronger statement was proven in \cite{\Wang}, namely
that $\W_{1 + \infty}$ at $c=-1$ can be written as
$\widehat{U(1)} \oplus \W(2,3)$ with the $\W(2,3)$ \cite{\Zam} at
$c=-2$. Since this algebra is a $\W(1,2,3)$ at $c=-1$, the result
of \cite{\Wang} implies ours.
The converse, {\it i.e.}\ that a $\widehat{U(1)}$ can be factored
out without affecting the field content of the remainder is a
special property of $\W_{1 + \infty}$ which is true for all $c$
(see e.g.\ \cite{\BEHHH}).

More generally, it has been conjectured in \cite{\BEHHH} that
$\W_{1+\infty}$ at $c=-L$ truncates to a $\W(1,2,\ldots,(L+1)^2-1)$.
This conjecture is based
on and supported by results of \cite{\AFMO}. Since it was shown
in \cite{\KaRa} that $\W_{1 + \infty}$ at $c=-L$ can be identified
with the $U(L)$-invariant vertex subalgebra of $L$ copies of the bosonic
ghost system, we expect that this conjecture can be proven
by generalizing the present computation to $U(L)$-invariants.
More specifically, one would need to derive a generalization of
\ref{begaFinRes} using the relations (VI) on p.\
71 of \cite{\Weyl} of which \ref{U1rel} is the $L=1$ special case.

At $c=2$ ($\lambda = 0$) we can consider the $U(1)$-invariant
vertex subalgebra generated by $\beta$ and $\der \gamma$. Now we
can eliminate all $U^{0,l}$ with $l \ge 5$ using \ref{begaFinRes}
and thus obtain a $\W(2,3,4,5)$ at $c=2$ generated by $U^{0,n}$
with $n\in \{1,2,3,4\}$. This algebra is actually the $c=2$
special case of a deformable $\W(2,3,4,5)$ that was first constructed by
direct methods in \cite{\Hornf}, as can be understood as follows:
The $\W(2,3,4,5)$ of \cite{\Hornf} can be realized in terms of
the coset $\widehat{sl(2,\Real)}_k/\widehat{U(1)}$ \cite{\BFH,\BEHHH}
and $c=2$ corresponds to the limit $k = \infty$. Using this
coset realization and general arguments along the lines of
\cite{\BFH} one can see that the limit $k \to \infty$ yields
precisely the $U(1)$-invariants of $J^{-}$ and $J^{+}/k$,
which have finite OPEs in the limit $k \to \infty$ and can then
be identified with $\beta$ and $\der \gamma$ using
the Wakimoto realization of $\widehat{sl(2,\Real)}_k$ (see
e.g.\ (15.279) in \cite{\dFMS}). In this manner also the
truncation of $\W_{\infty}$ at $c=2$ to a $\W(2,3,4,5)$
\cite{\BEHHHt} arises naturally, since our construction
clearly gives rise to a linear algebra in the basis of
the $U^{0,l}$. Actually, truncations of $\W_{\infty}$ are
expected for $c \in 2 \Zed$ \cite{\BEHHHt,\BEHHH} and one
may speculate that our result on the $U(1)$-invariants
generated by $\beta$ and $\der \gamma$ can be generalized
in a way similar to $\W_{1+\infty}$.

Finally,
recall that at the symmetric point $\lambda = {1 \over 2}$ ($c=-1$),
the symmetry is enlarged to an $Sp(2,\Complex)$. The invariants of
this group were studied in \cite{\BFH}. There it was shown by an
explicit computation of normal-ordered relations that the
$Sp(2,\Complex)$-invariant vertex subalgebra of the $\beta$-$\gamma$
system is of type $\W(2,4,6)$ \footnote{${}^{5})$}{This algebra is
the $c=-1$ special case of a $\W(2,4,6)$ constructed in \cite{\KaWa}
by direct methods. The identification at $c=-1$ and its generalizations
in \cite{\BFH} solved the problem \cite{\EHH} of identifying this
$\W(2,4,6)$-algebra.
}. The computation of \cite{\BFH}
differs from the present one slightly in that it used rearrangement
lemmas, but to the best of our knowledge this is the only non-trivial
case where an invariant vertex subalgebra was rigorously shown to
be of a certain type (viewed as a $\W$-algebra) prior to the present
work.
\bn
\section{Conclusion}
\mn
The main results of this paper are summarized as follows:

In the first part we formulated fermionic
and bosonic ghost systems as a single vertex algebra each. These
vertex algebras admit the one-parameter families \ref{stateEMTbc}
and \ref{stateEMTbega} of conformal vectors. We observed the
simple relation \ref{EMTbcZero} between the actions of the Virasoro
algebra in each family. A corollary is the identity \ref{bcCharId}
between characters for different choices of conformal vectors.
With this information one readily obtains the explicit formulae
\ref{bcChar} and \ref{begaChar} for the characters of a general
twisted module of a ghost system.

The second part was devoted to symmetries and vertex subalgebras
invariant under them. First, we have discussed the invariant theory
for the group $\Zed_N$, including previous results \cite{\BFH} for
$\Zed_2$ and $U(1)$ as limiting cases. Then we have studied
the $\Zed_N$-invariant subalgebras of the fermionic ghost system in
detail and shown that they are $\W$-algebras of type
$\W(1,{N^2 \over 2} + (\lambda - {1 \over 2}) N,
{N^2 \over 2} + ({1 \over 2} - \lambda) N)$. At $c=-2$ the
$\Zed_N$-invariant subalgebras based on $b$ and $\der c$
({\it i.e.}\ omitting $c$) are conjectured to be of type
$\W(2,3,{N (N+1) \over 2}, {N (N+1) \over 2})$. However this
has been shown rigorously only for $N=2$ and $N = \infty$
where we recover a $\W(2,3^3)$ \cite{\multiplet,\Kausch} and a
$\W(2,3)$ \cite{\Zam} at $c=-2$, respectively. Finally,
we showed that the $U(1)$-invariant subalgebra of the
generic bosonic ghost system is an algebra of type
$\W(1,2,3)$. For $\lambda=0$ the $U(1)$-invariant subalgebra of
$\beta$ and $\der \gamma$ is a $\W(2,3,4,5)$ \cite{\Hornf} with $c=2$.

We should mention that the ghost-number currents \ref{bcCurrent}
and \ref{begaCurrent} are primary only at the symmetric point
$\lambda={1 \over 2}$. This means that if (as is sometimes done)
we require the generators of a $\W$-algebra to be primary
fields, we have to specialize the
$\W(1,{N^2 \over 2} + (\lambda - {1 \over 2}) N,
{N^2 \over 2} + ({1 \over 2} - \lambda) N)$ and the
$\W(1,2,3)$ obtained from the fermionic and bosonic ghost
system, respectively to this symmetric point ($c=1$ and $c=-1$,
respectively).

The question of (generalized) rationality of the conformal field
theories associated to the algebras studied in this paper remains
to be investigated. This will involve logarithmic fields
\cite{\Gurarie} since they are known to appear in certain
rational conformal field theories at $c= -2$ \cite{\Flohr,\GaKa}.

In the present paper we have always worked with a single ghost system.
The discussion of the first part generalizes to several
copies with arbitrarily chosen energy-momentum tensor and twist in
each copy in the obvious way. A generalization of our study of invariant
vertex subalgebras to several copies and other groups is more
difficult, but would be very interesting. One application would
be to check the conjecture of \cite{\BEHHH} on the structure of
$\W_{1+\infty}$ at $c=-L$ using its construction in terms of the
$U(L)$-invariants in $L$ copies of the $c=-1$ bosonic ghost system
\cite{\KaRa}. More generally, in this way one may hope to gain further
insight into the structure of conformal field theories at integer
central charge. 
\vskip 1.5 cm
\displayhead{Acknowledgments}
\mn
L.F.\ is grateful to the Dublin Institute for Advanced Studies
for hospitality during the course of this work.
\sectionnumstyle{Alphabetic}
\newsectionnum=0
\vskip 4.0 cm
\appendix{Twisted modules \& useful formulae}
\mn
In this appendix we summarize the main ingredients of the definition of
twisted modules of vertex algebras \cite{\KaRa} and add the
necessary signs to include fermions. We also collect some useful
formulae for the computations performed in this paper.

A twisted vertex operator $Y$ assigns a twisted field $Y(\state{a},z)$ 
to any state $\state{a}$ of a vertex algebra. When $Y$ is the vacuum 
vertex operator the twist is trivial and this map becomes 
an isomorphism which also plays a central r\^ole in conformal field 
theory (see e.g.\ \cite{\Goddard}). 
Throughout the remainder of this paper we use the notation
$$f(z) = Y(\state{f},z) = \sum_{n\in \phi+\Zed} f_n^{(\alpha)} z^{-n-1}\, ,
\label{vertexMap}$$
where $\state{f} = f_{-1}\vac$ is a state in the vacuum module of the vertex
algebra  and $f(z)$ is a $\phi$-twisted field; only for the purposes of the 
appendix we keep the vertex map $Y$ explicitly. Here the modes of a 
twisted field are denoted similarly to the bulk of the paper by an
$\alpha$ as upper index characterizing the twisted module
(which we omit for the vacuum module).

A twisted module is then characterized by the following three axioms 
\cite{\KaRa}:
{\parindent=10mm
\item{(M1)} $Y(\vac,z) = \id^{(\alpha)}$,
\item{(M2)} $Y(L_{-1} \state{a},z) = \der Y(\state{a},z)$,
\item{(M3)} (twisted Borcherds identity)
$$\eqalign{
\sum_{j=0}^{\infty} {m \choose j} Y(a_{n+j} \state{b}, z) z^{m-j}
=& \sum_{j=0}^{\infty} (-1)^j {n \choose j} \left\{
a^{(\alpha)}_{m+n-j} Y(\state{b}, z) z^j \right. \cr
&\quad \left. - (-1)^n \epsilon_{a,b} Y(\state{b}, z)
   a^{(\alpha)}_{m+j} z^{n-j}
\right\} \, ,
}\label{twistBorId}$$
\item{}
where $n \in \Zed$, $m \in \Zed + \alpha$ and $\alpha$ is the twist
of the field $a(z)$. Here $\epsilon_{a,b} = -1$ if both $a$ and $b$
are fermions and $\epsilon_{a,b} = 1$ otherwise. \par
}

The $m= \alpha$, $n=-1$ special case of \ref{twistBorId} yields the
following formula for normal-ordered products in a twisted module
\cite{\KaRa}:
$$\sum_{j=0}^{\infty} {\alpha \choose j} Y(a_{j-1} \state{b},z) z^{-j}
= \;: Y(\state{a},z) Y(\state{b},z) :
  \, ,
\label{twistNOP}$$
where the normal-ordered product of the two fields
$a$ and $b$ is defined with the mode-expansion \ref{modConv} by
$$: Y(\state{a},z) Y(\state{b},z) :_n =
    \sum_{k < -\alpha} a_k b_{n-1-k} + \epsilon_{a,b}
                 \sum_{k \ge -\alpha} b_{n-1-k} a_k \, .
\label{NOPcon}$$
Here as in (M3) $\epsilon_{a,b} = 1$ unless both $a$ and $b$ are
fermions, and $\alpha$ is the twist of $a$.

It should be noted that one usually also requires that for all 
$\state{v}$ in the twisted module and all $\state{a}$ in the 
vertex algebra $a_n^{(\alpha)} \state{v}$ vanishes for $n$ large 
enough (in the cases we consider this is indeed true as we are 
only dealing with highest weight representations).
This ensures that the sums on the l.h.s.\ of \ref{twistBorId} and also 
\ref{twistNOP} are actually finite. In fact, for our purposes the 
sums are truncated in a manner that at most three terms in the vacuum
representation of the ghost systems need to be considered.

We conclude this axiomatic part by mentioning that a vertex algebra
is called simple if its vacuum representation is irreducible.

Finally, we collect a few useful formulae. Without a twist
one has the highest-weight condition for the
vacuum vector $\vac$: $\phi_k \vac = 0$ for all $k \ge 0$.
Combining this with the specialization of \ref{NOPcon} to
a trivial twist implies
$\NOP{a}{b}_{-1} \vac = a_{-1} b_{-1} \vac$.

{}From the mode expansion \ref{modConv} one finds furthermore
$$(\der^n \phi)_k = (-1)^n k (k-1) \cdots (k-n+1) \phi_{k-n}
\, .
\label{MODEder}$$
Finally, the Virasoro-generators $L^{(\alpha)}_m$ act as
$$[L^{(\alpha)}_m, \phi^{(\alpha)}_n]
   = \left( (\dim(\phi) -1) (m+1) -n \right) \phi^{(\alpha)}_{m+n} \, ,
\label{virCom}$$
which is valid 
\item{$\bullet$} for all $m$ if $\phi$ is a {\it primary} field of
                conformal dimension $\dim(\phi)$,
\item{$\bullet$} for $m=0$ and $m=\pm 1$ if $\phi$ is
                a {\it quasi-primary} field of
                conformal dimension $\dim(\phi)$,
\item{$\bullet$} for $m=0$ and $m=- 1$ if $\phi$ has a definite
                conformal dimension $\dim(\phi)$, and
\item{$\bullet$} for $m=-1$ always.

In particular, from combination of \ref{MODEder} with \ref{virCom}
one recovers the axiom (M2).
\vfill
\eject
\appendix{Orbifold characters}
\mn
In this appendix we discuss the characters of the $\Zed_N$-orbifolds
of the fermionic ghost system at $c=1$ ({\it i.e.}\ the complex fermion).

For this end we need a suitable form of the characters of the
fermionic ghost system with a twist $\alpha$. One reads off
from \ref{bcChar} that $\chi^{(\lambda,\alpha)}(q,z) = z^{\alpha}
\chi^{(\lambda+\alpha,0)}(q,z)$. Inserting this into
\ref{bcCharSumFgen} one finds
$$\chi^{({1 \over 2},\alpha)}(q,z) = {z^{\alpha} q^{{\alpha^2 \over 2}}
\over \eta(q)} \sum_{n \in \Zed} q^{{n^2 \over 2} + n \alpha} z^n \, .
\label{bcCharSumTgen}$$
Passing now to the $\Zed_N$-orbifolds, each module of the ghost
system with twist $N \alpha \in \Zed$ will split into untwisted
modules of the orbifold algebra. The characters of the modules
of the orbifold are just the coefficient of $\omega^n$ in
$\chi^{({1 \over 2},\alpha)}(q,\omega)/\omega^{\alpha}$ (where $\omega$
is the primitive $N$th root of unity). There are $N$ twisted modules,
each splitting into $N$ modules for the orbifold and therefore the
orbifold algebra has a total of $N^2$ characters.
One obtains from \ref{bcCharSumTgen} and with a few manipulations the
following orbifold-characters
$$\chi_s(q) 
= {q^{s^2 \over 2 N^2} \over \eta(q)} \sum_{n \in \zed}
    q^{{n^2 N^2 \over 2} + n s}
= {1 \over \eta(q)} \sum_{n\in\zed} q^{(s+nN^2)^2\over 2N^2}
 \, ,
\label{bcZnOrbChar}$$
where $0 \le s \le N^2-1$. This yields the partition function
$$Z(q) = \sum_{s=0}^{N^2-1} \chi_s(\qb) \chi_s(q)
= {1 \over \eta(\qb) \eta(q)} \sum_{{\kb,k \in \zed} \atop
    {\kb-k \in N^2 \zed}} \qb^{{\kb^2 \over 2 N^2}} q^{{k^2 \over 2 N^2}} \, .
\label{bcZnPartF}$$
{}From this one immediately recognizes the partition function of a free
boson compactified at radius $r={N \over 2}$ if $N$ is even
(in the notation of \cite{\Gin}). This identification is also valid
for odd $N$, with the only further subtlety that the Ramond sector of the
orbifold algebra must be added to \ref{bcZnPartF} to recover the full
modular invariant partition function of the free boson at
radius $r={N \over 2}$.
\vfill
\eject
\displayhead{References}
\mn
\bibitem{\FMS} D.\ Friedan, E.\ Martinec, S.\ Shenker, {\it Conformal
              Invariance, Supersymmetry and String Theory}, Nucl.\ Phys.\
              {\bf B271} (1986) 93-165
\bibitem{\GuLu} S.\ Guruswamy, A.W.W.\ Ludwig, {\it Relating $c<0$ and $c>0$
              Conformal Field Theories}, preprint hep-th/9612172, IC/96/273
\bibitem{\GeTo} L.S.\ Georgiev, I.T.\ Todorov, {\it Characters and Partition
              Function for the Wen--Wu CFT Model of the Haldane--Rezayi
              Quantum Hall State}, preprint hep-th/9611084, INRNE-TH-96/13
\bibitem{\FGN} V.\ Gurarie, M.\ Flohr, C.\ Nayak, {\it The Haldane-Rezayi
              Quantum Hall State and Conformal Field Theory},
              Nucl.\ Phys.\ {\bf B498} (1997) 513-538, {\tt cond-mat/9701212}
\bibitem{\MoRe} G.\ Moore, N.\ Read, {\it Nonabelions in the Fractional Quantum
              Hall Effect}, Nucl.\ Phys.\ {\bf B360} (1991) 362-396
\bibitem{\Saleur} H.\ Saleur, {\it Polymers and Percolation in Two Dimensions
              and Twisted $N=2$ Supersymmetry}, Nucl.\ Phys.\ {\bf B382} (1992)
              486-531
\bibitem{\Bernard} D.\ Bernard, {\it (Perturbed) Conformal Field Theory Applied
              to 2D Disordered Systems: An Introduction}, preprint
              hep-th/9509137, SPhT-95-113, IHES/P/95/85
\bibitem{\MaSe} Z.\ Maassarani, D.\ Serban, {\it Non-Unitary Conformal Field
              Theory and Logarithmic Operators for Disordered Systems}, Nucl.\
              Phys.\ {\bf B489} (1997) 603-625, {\tt hep-th/9605062}
\bibitem{\dFMS} P.\ di Francesco, P.\ Mathieu, D.\ S\'en\'echal,
              {\it Conformal Field Theory}, Graduate Texts in Contemporary
              Physics, Springer-Verlag, New York (1997) 
\bibitem{\JdBLF} J.\ de Boer, L.\ Feh\'er, {\it Wakimoto Realizations of Current
              Algebras: An Explicit Construction}, preprint hep-th/9611083,
              LBNL-39562, UCB-PTH-96/49, BONN-TH-96/16
\bibitem{\Ras} J.\ Rasmussen, {\it Free Field Realizations of Affine Current
              Superalgebras, Screening Currents and Primary Fields}, preprint
              hep-th/9706091, NBI-HE-97-15
\bibitem{\Kausch} H.G.\ Kausch, {\it Curiosities at $c=-2$}, preprint
              hep-th/9510149, DAMTP 95-52
\bibitem{\Gin} P.\ Ginsparg, {\it Curiosities at $c=1$}, Nucl.\ Phys.\ {\bf
              B295} (1988) 153-170
\bibitem{\Kir} E.B.\ Kiritsis, {\it Proof of the Completeness of the
              Classification of Rational Conformal Theories with $c=1$},
              Phys.\ Lett.\ {\bf B217} (1989) 427-430
\bibitem{\Borcherds} R.E.\ Borcherds, {\it Vertex Algebras, Kac-Moody Algebras,
              and the Monster}, Proc.\ Natl.\ Acad.\ Sci.\ USA {\bf 83} (1986)
              3068-3071
\bibitem{\KaRa} V.\ Kac, A.\ Radul, {\it Representation Theory of the Vertex
              Algebra $\W_{1+\infty}$}, Transform.\ Groups {\bf 1} (1996) 41-70,
              {\tt hep-th/9512150}
\bibitem{\BEHHH} R.\ Blumenhagen, W.\ Eholzer, A.\ Honecker, K.\ Hornfeck, R.\
              H\"ubel, {\it Coset Realization of Unifying $\W$-Algebras},
              Int.\ Jour.\ of Mod.\ Phys.\ {\bf A10} (1995) 2367-2430,
              {\tt hep-th/9406203}
\bibitem{\FRRTW} L.\ Feh\'er, L.\ O'Raifeartaigh, P.\ Ruelle, I.\ Tsutsui, A.\
              Wipf, {\it On Hamiltonian Reductions of the
              Wess-Zumino-Novikov-Witten Theories}, Phys.\ Rep.\ {\bf 222}
              (1992) 1-64
\bibitem{\FKW} E.\ Frenkel, V.\ Kac, M.\ Wakimoto, {\it Characters and
              Fusion Rules for $\W$-Algebras via Quantized Drinfeld-Sokolov
              Reduction}, Commun.\ Math.\ Phys.\ {\bf 147} (1992) 295-328
\bibitem{\dBT} J.\ de Boer, T.\ Tjin, {\it The Relation between Quantum $\W$
              Algebras and Lie Algebras}, Commun.\ Math.\ Phys.\ {\bf 160}
              (1994) 317-332, {\tt hep-th/9302006}
\bibitem{\Schell} A.N.\ Schellekens, {\it Meromorphic $c=24$ Conformal Field
              Theories}, Commun.\ Math.\ Phys.\ {\bf 153} (1993) 159-185,
              {\tt hep-th/9205072}
\bibitem{\CaTrZe} A.\ Cappelli, C.A.\ Trugenberger, G.R.\ Zemba, {\it
              Classification of Quantum Hall Universality Classes by
              $\W_{1+\infty}$ Symmetry}, Phys.\ Rev.\ Lett.\ {\bf 72} (1994)
              1902-1905, {\tt hep-th/9310181}
\bibitem{\FlVa} M.\ Flohr, R.\ Varnhagen, {\it Infinite Symmetry in the
              Fractional Quantum Hall Effect}, J.\ Phys.\ A: Math.\ Gen.\
              {\bf 27} (1994) 3999-4010, {\tt hep-th/9309083}
\bibitem{\KacRad} V.\ Kac, A.\ Radul, {\it Quasifinite Highest Weight Modules
              over the Lie Algebra of Differential Operators on the Circle},
              Commun.\ Math.\ Phys.\ {\bf 157} (1993) 429-457, {\tt
              hep-th/9308153}
\bibitem{\FKRW} E.\ Frenkel, V.\ Kac, A.\ Radul, W.\ Wang, {\it
              $\W_{1+\infty}$ and $\W(gl_N)$ with Central Charge $N$},
              Commun.\ Math.\ Phys.\ {\bf 170} (1995) 337-357,
              {\tt hep-th/9405121}
\bibitem{\AFMO} H.\ Awata, M.\ Fukuma, Y.\ Matsuo, S.\ Odake, {\it Character
              and Determinant Formulae of Quasifinite Representation of the
              $\W_{1+\infty}$ Algebra}, Commun.\ Math.\ Phys.\ {\bf 172}
              (1995) 377-400, {\tt hep-th/9405093}
\bibitem{\KaTo} V.G.\ Kac, I.T.\ Todorov, {\it Affine Orbifolds and Rational
              Conformal Field Theory Extensions of $\W_{1+\infty}$}, preprint
              hep-th/9612078
\bibitem{\Kac} V.\ Kac, {\it Vertex Algebras for Beginners}, University
              Lecture Series Vol.\ 10, American Mathematical Society,
              Providence, R.I.\ (1997)
\bibitem{\KacRaina} V.G.\ Kac, A.K.\ Raina, {\it
              Bombay Lectures on Highest Weight Representations of
              Infinite Dimensional Lie Algebras}.
              Advanced Series in Mathematical Physics Vol.\ 2,
              World Scientific, Singapore (1987)
\bibitem{\SchS} A.\ Schwimmer, N.\ Seiberg, {\it Comments on the $N=2$, $3$,
              $4$ Superconformal Algebras in Two Dimensions}, Phys.\ Lett.\
              {\bf B184} (1987) 191-196
\bibitem{\FeiFr} B.\ Feigin, E.\ Frenkel, {\it Bosonic Ghost System and the
              Virasoro Algebra}, Phys.\ Lett.\ {\bf B246} (1990) 71-74
\bibitem{\Auto} A.\ Honecker, {\it Automorphisms of $\W$-Algebras and Extended
              Rational Conformal Field Theories}, Nucl.\ Phys.\ {\bf B400}
              (1993) 574-596, {\tt hep-th/9211130}
\bibitem{\Weyl} H.\ Weyl, {\it The Classical Groups, Their Invariants
              and Representations}, Princeton University Press,
              Princeton, New Jersey (1946)
\bibitem{\BFH} J.\ de Boer, L.\ Feh\'er, A.\ Honecker, {\it A Class of
              $\W$-Algebras with Infinitely Generated Classical Limit}, Nucl.\
              Phys.\ {\bf B420} (1994) 409-445, {\tt hep-th/9312049}
\bibitem{\GinProc} P.\ Ginsparg, {\it Applied Conformal Field Theory}, 
              pp.\ 1-168 in: E.\ Br\'ezin, J.\ Zinn-Justin (eds.),
              {\it Fields, Strings and Critical Phenomena},
              Proceedings of the Les Houches Summer School 1988,
              North-Holland, Amsterdam (1990)
\bibitem{\Zam} A.B.\ Zamolodchikov, {\it Infinite Additional Symmetries in
              Two-Dimensional Conformal Quantum Field Theory}, Theor.\ Math.\
              Phys.\ {\bf 65} (1986) 1205-1213
\bibitem{\BCNMcC} P.\ Bouwknegt, A.\ Ceresole, P.\ van Nieuwenhuizen,
              J.\ McCarthy, {\it Extended Sugawara Construction for the
              Superalgebra $SU(M+1|N+1)$.\ II.\ The Third-Order Casimir
              Algebra}, Phys.\ Rev.\ {\bf D40} (1989) 415-421
\bibitem{\multiplet} H.G.\ Kausch, {\it Extended Conformal Algebras Generated
              by a Multiplet of Primary Fields}, Phys.\ Lett.\ {\bf B259}
              (1991) 448-455
\bibitem{\EHH} W.\ Eholzer, A.\ Honecker, R.\ H\"ubel, {\it How Complete is
              the Classification of $\W$-Symmetries?}, Phys.\ Lett.\ {\bf
              B308} (1993) 42-50, {\tt hep-th/9302124}
\bibitem{\BEHHHt} R.\ Blumenhagen, W.\ Eholzer, A.\ Honecker, K.\ Hornfeck, R.\
              H\"ubel, {\it Unifying $\W$-Algebras}, Phys.\ Lett.\ {\bf B332}
              (1994) 51-60, {\tt hep-th/9404113}
\bibitem{\Wang} W.\ Wang, {\it $\W_{1+ \infty}$ Algebra, $\W_3$ Algebra, and
              Friedan-Martinec-Shenker Bosonization}, preprint q-alg/9708008
\bibitem{\Hornf} K.\ Hornfeck, {\it $\W$-Algebras with Set of Primary Fields of
              Dimensions (3,4,5) and (3,4,5,6)}, Nucl.\ Phys.\ {\bf B407}
              (1993) 237-246, {\tt hep-th/9212104}
\bibitem{\KaWa} H.G.\ Kausch, G.M.T.\ Watts, {\it A Study of $\W$-Algebras
              Using Jacobi Identities}, Nucl.\ Phys.\ {\bf B354} (1991)
              740-768
\bibitem{\Gurarie} V.\ Gurarie, {\it Logarithmic Operators in Conformal Field
              Theory}, Nucl.\ Phys.\ {\bf B410} (1993) 535-549,
              {\tt hep-th/9303160}
\bibitem{\Flohr} M.A.I.\ Flohr, {\it On Fusion Rules in Logarithmic Conformal
              Field Theories}, Int.\ Jour.\ of Mod.\ Phys.\ {\bf A12} (1997)
              1943-1958, {\tt hep-th/9605151}
\bibitem{\GaKa} M.R.\ Gaberdiel, H.G.\ Kausch, {\it A Rational Logarithmic
              Conformal Field Theory}, Phys.\ Lett.\ {\bf B386} (1996)
              131-137, {\tt hep-th/9606050}
\bibitem{\Goddard} P.\ Goddard, {\it Meromorphic Conformal Field Theory},
              pp.\ 556-587 in: V.G.\ Kac (ed.),
              {\it Infinite Dimensional Lie Algebras and Groups},
              Proceedings of the CIRM-Luminy Conference 1988,
              World Scientific, Singapore (1989)

\vfill
\end